\documentclass[10pt, prd, twocolumn, nofootinbib,preprint,superscriptaddress]{revtex4}
\pdfoutput=1
\usepackage[T1]{fontenc}
\usepackage{comment}
\usepackage{amsmath,amssymb,ulem}
\usepackage{epsfig}
\usepackage{graphicx}
\usepackage[usenames,dvipsnames]{color}
\usepackage{subfigure}
\usepackage{slashed}
\usepackage[colorlinks,citecolor=blue]{hyperref}
\usepackage{pdfpages}
\usepackage{color}

\begin{document}
\title{A first order dark $SU(2)_D$ phase transition with vector dark matter \\ in the light of NANOGrav 12.5 yr data}

\author{Debasish Borah}
\email{dborah@iitg.ac.in}
\affiliation{Department of Physics, Indian Institute of Technology Guwahati, Assam 781039, India}
\author{Arnab Dasgupta}
\email{arnabdasgupta@pitt.edu}
\affiliation{Pittsburgh Particle Physics, Astrophysics, and Cosmology Center, Department of Physics and Astronomy, University of Pittsburgh, Pittsburgh, PA 15206, USA}
\affiliation{Institute of Convergence Fundamental Studies , Seoul-Tech, Seoul 139-743, Korea}
\author{Sin Kyu Kang}
\email{skkang@seoultech.ac.kr}
\affiliation{School of Liberal Arts, Seoul-Tech, Seoul 139-743, Korea}

\begin{abstract}
We study a dark $SU(2)_D$ gauge extension of the standard model (SM) with the possibility of a strong first order phase transition (FOPT) taking place below the electroweak scale in the light of NANOGrav 12.5 yr data. As pointed out recently by the NANOGrav collaboration, gravitational waves (GW) from such a FOPT with appropriate strength and nucleation temperature can explain their 12.5 yr data. We impose a classical conformal invariance on the scalar potential of $SU(2)_D$ sector involving only a complex scalar doublet with negligible couplings with the SM Higgs. While a FOPT at sub-GeV temperatures can give rise to stochastic GW around nano-Hz frequencies being in agreement with NANOGrav findings, the $SU(2)_D$ vector bosons which acquire masses as a result of the FOPT in dark sector, can also serve as dark matter (DM) in the universe. The relic abundance of such vector DM can be generated in a non-thermal manner from the SM bath via scalar portal mixing. We also discuss future sensitivity of gravitational wave experiments to the model parameter space.
\end{abstract}

\maketitle

\noindent
{\bf Introduction}: The NANOGrav collaboration has recently searched for a gravitational wave (GW) signal produced from a first order phase transition (FOPT) in 45 pulsars data set spanning over 12.5 yr \cite{Arzoumanian:2021teu}. According to their analysis, the 12.5 yr data can be interpreted in terms of a FOPT occurring at a temperature below the electroweak (EW) scale.  It should however be noted that similar effects at the NANOGrav experiment can also be caused by signals generated by supermassive black hole binary (SMBHB) mergers. In 2020, the NANOGrav collaboration also reported that they have found strong evidence of a stochastic process, modeled as a power-law spectrum having frequencies around the nano-Hz regime, with common amplitude
and spectral slope across pulsars, in their 12.5 yr data \cite{Arzoumanian:2020vkk}. Although the statistical significance for a stochastic GW background discovery claim was low, it still led to several interesting new physics explanations like cosmic string origins \cite{Blasi:2020mfx, Ellis:2020ena, Bian:2021lmz}, FOPT origins \cite{Ratzinger:2020koh, Addazi:2020zcj, Nakai:2020oit, Bian:2021lmz, Zhou:2021cfu, Borah:2021ocu}, inflationary origin \cite{Vagnozzi:2020gtf} etc. While more data are required to settle these issues, the pulsar timing arrays (PTAs) like NANOGrav sensitive to GW of extremely low frequencies do offer a complementary probe of GW background to future space-based interferometers like eLISA \cite{Caprini:2015zlo,Caprini:2019egz}. Future data have the potential to probe many of the above-mentioned new physics explanations for such low frequency GW background. As pointed out recently, even different experiments like GAIA and its future upgrades have the potential to probe NANOGrav results \cite{Garcia-Bellido:2021zgu}. Interestingly, the recent results from the PPTA collaboration, another PTA based GW experiment, have also indicated similar findings consistent with NANOGrav observations \cite{Goncharov:2021oub}. 

Motivated by the recent results from NANOGrav collaboration explained in terms of a FOPT characterized by the preferred ranges for strength $(\alpha_*)$ as well as the phase transition temperature $(T_*)$ as shown in \cite{Arzoumanian:2021teu},
we study a simple model to achieve such a strong FOPT below electroweak scale. The idea of such strong FOPT at low scale has been explored within the context of Abelian gauge extended models with or without additional dark matter (DM) fields charged under the same gauge symmetry. More specifically, in our recent work \cite{Borah:2021ocu}, a complex scalar singlet charged under such Abelian gauge symmetry while simultaneously obeying classical conformal invariance led to a strong FOPT at bubble nucleation temperature much below electroweak scale. For earlier works on FOPT within such Abelian gauge extended scenarios, please refer to \cite{Jinno:2016knw, Mohamadnejad:2019vzg, Kim:2019ogz, Hasegawa:2019amx, Marzo:2018nov, Hashino:2018zsi, Chiang:2017zbz} and references therein. In this work, we consider the possibility of realising such a low scale FOPT within a dark non-Abelian gauge model, particularly focusing on a $SU(2)_D$ gauge extension. Several earlier works on such dark FOPT with non-Abelian gauge symmetries have already studied the consequences for GW at eLISA type experiments \cite{Schwaller:2015tja, Baldes:2018emh, Prokopec:2018tnq}. On the other hand, in this work we focus on the possibility of having a FOPT within such non-Abelian dark sectors at sub-EW scale such that the resulting stochastic GWs can have frequencies typically within the NANOGrav or other present as well as future PTA type experiments. 

While such dark strong FOPT and resulting GWs have been investigated earlier as well, we study this possibility within a dark non-Abelian gauge sector for the first time after NANOGrav collaboration analysed their 12.5 year data in the context of GWs from the FOPT at a sub-EW scale \cite{Arzoumanian:2021teu}. Another motivation for such dark gauge symmetries like $U(1)_D$, as discussed in earlier work \cite{Borah:2021ocu}, is that it can also be motivated from DM point of view where a singlet field charged under this gauge symmetry plays the role of DM. On the other hand, a dark non-Abelian gauge symmetry like $SU(2)_D$, as we adopt in this work, naturally contains a DM candidate in the form of the massive vector boson\footnote{Note that $U(1)_D$ vector boson can also be made a viable DM candidate by tuning the kinetic mixing $U(1)_D$ and $U(1)_Y$ of standard model, making it cosmologically long-lived.}. We consider such a minimal scenario of $SU(2)_D$ gauge symmetry with a complex scalar field responsible for symmetry breaking and constrain the parameter space from the requirement of providing one possible explanation of NANOGrav observations along with observed DM relic density from Planck \cite{Aghanim:2018eyx}. We also show the future sensitivity of GW experiments \cite{Schmitz:2020syl} like SKA \cite{Weltman:2018zrl}, IPTA \cite{Hobbs_2010} to the model parameter space. \\

\noindent
{\bf The Model}: As demonstrated above, we study a $SU(2)_D$ extension of the standard model (SM). The newly introduced field in this model is a complex scalar doublet $\Phi$ required for spontaneous breaking of gauge symmetry. All the SM fields are neutral under this new gauge symmetry. The zero-temperature scalar potential at tree level is given by
\begin{align}
    V_{\rm tree} = \lambda_1 (H^{\dagger} H)^2 + \lambda_2 (\Phi^{\dagger} \Phi) (H^{\dagger} H) +\lambda_3 ( \Phi^{\dagger} \Phi)^2,
    \label{scalpot1}
\end{align}
where $H$ is the SM Higgs doublet. As we impose classical conformal invariance, the scalar potential remains free from bare mass terms. The kinetic terms for the $SU(2)_D$ sector fields can be written as
\begin{align}
    \mathcal{L}_{\rm kin} \supset (D_{\mu} \Phi)^{\dagger} (D^{\mu} \Phi)-\frac{1}{4} (F_D)_{\mu \nu} (F_D)^{\mu \nu},
\end{align}
where $F_D$ is the field strength tensor for $SU(2)_D$ and $D_{\mu} \Phi = (\partial_{\mu}+ig_D \tau^i \cdot A^i_D/2) \Phi$ is the covariant derivative. The dark scalar doublet $\Phi$ can be written in component form as
\begin{equation}
\Phi=\frac{1}{\sqrt{2}}\begin{pmatrix} 
  G_2 + i G_3\\ 
  M + \phi + i G_1
\end{pmatrix}
\end{equation}

The vacuum expectation value (VEV) of the dark scalar doublet, $\langle \Phi \rangle=M/\sqrt{2}$, acquired via the running of the quartic coupling $\lambda_3$ breaks the $SU(2)_D$ gauge symmetry leading to a massive gauge boson $M_{Z_D}=g_D M/2$. In order to realize the EW vacuum, the coupling $\lambda_2$ needs to be suppressed. Therefore, in our analysis we neglect the coupling $\lambda_2$. This assumption is for simplicity and also to make sure that the SM Higgs VEV does not affect the light singlet scalar mass. As we will see later, this assumption also helps in ensuring non-thermal production of light vector boson DM via scalar portal. Therefore, the suppressed coupling of $SU(2)_D$ sector particles with the SM guarantees the latter's contribution to be negligible as to suppress its role in the renormalisation group evolution (RGE) of the singlet scalar quartic coupling.

The total effective potential is schematically composed of the following three terms:
\begin{align}
V_{\rm tot} = V_{\rm tree} + V_{\rm CW} +V_T,
\end{align}
where $V_{\rm tree},~V_{\rm CW}$ and $V_T$ denote the tree level scalar potential, the one-loop Coleman-Weinberg potential, and the thermal effective potential, respectively. In finite-temperature field theory, the effective potential, $V_{\rm CW}$ and $V_T$, are calculated by using the standard background field method~\cite{Dolan:1973qd,Quiros:1999jp}. While we assume Landau gauge in our analysis, issues related to gauge dependence in such conformal models can be found in \cite{Chiang:2017zbz}. Denoting the dark scalar doublet as above, the zero temperature effective potential is given as
\begin{align}
     V_{0} &= V_{\rm tree} + V_{\rm CW}, 
\end{align}
where $ V_{\rm tree}= \lambda_3 \phi^4/4$ and $V_{\rm CW}$ is given by
\begin{equation}
    V_{\rm CW} = \frac{1}{64 \pi^2} \sum_i (2s_i+1) m^4_i (\phi) \bigg [ \ln{ \left( \frac{m^4_i(\phi)}{\mu^2} \right )}-C_i  \bigg ].
\end{equation}
Here $s_i$ is the spin with the index $i$ runs over gauge boson and scalars. Their field dependent masses are
\begin{equation}
    m^2_D (\phi) = g^2_D \phi^2/2, m^2_{\phi}= 3 \lambda_3 \phi^2, m^2_G=\lambda_3 \phi^2.
\end{equation}
In the expression for $V_{\rm CW}$, the constant $C_i=5/2$ for gauge bosons and $C_i=3/2$ otherwise. The $\overline{\rm MS}$ renormalisation scale is denoted by $\mu$.

The gauge coupling $g_D(t)$ and 
quartic coupling $\lambda_3(t)$ at renormalisation scale can be calculated by 
solving the corresponding RGE equations. In 
terms of $\alpha_D=g^2_D/4\pi$ and $\alpha_{\lambda}=\lambda_3/4\pi$, the RGEs are
\begin{align}
    \frac{d\alpha_D(t)}{dt}=\frac{b}{2\pi} \alpha^2_D(t),
\end{align}
\begin{align}
    \frac{d\alpha_{\lambda}(t)}{dt}=\frac{1}{2\pi} \left( a_1 \alpha^2_{\lambda}(t)+8\pi \alpha_{\lambda}(t) \gamma(t)+a_3 \alpha^2_D(t) \right),
\end{align}
where $t=\ln{(\phi/\mu)}$ and, $\gamma (t) =-a_2 \alpha_D(t)/(8\pi) $. For $SU(2)_D$ gauge group, $b=-43/3$, $a_1=24, a_2=9/2$, and $a_3=9/16$. 
Taking the renormalisation scale $\mu$ to be $M$, the condition $\frac{dV}{d\phi}|_{\phi=M}=0$ leads us to the relation,
\begin{align}
    a_1 \alpha_\lambda^2(0) + a_3\alpha^2_{D}(0) + 8\pi \alpha_\lambda(0) &=0,
    \label{eq:eq1}
\end{align}
from which $\alpha_\lambda(0)$ is determined in terms of $\alpha_{D}(0)$.
Using the analytic solutions of the above RGEs, the scalar potential can be written as \cite{Jinno:2016knw}
\begin{align}
    V_{0}(\phi,t) &= \frac{\pi \alpha_\lambda(t)}{(1-\frac{b}{2\pi}\alpha_{D}(0)t)^{a_2/b}}\phi^4
\end{align}
where
\begin{align}
    \alpha_{D}(t) &= \frac{\alpha_{D}(0)}{1-\frac{b}{2\pi}\alpha_{D}(0)t} \\
    \alpha_\lambda(t) &= \frac{a_2+b}{2a_1}\alpha_{D}(t) \nonumber \\
    &+ \frac{A}{a_1}\alpha_{D}(t)\tan \left[\frac{A}{b}\ln[\alpha_{D}(t)/\pi] +C\right]  \nonumber \\
    A &\equiv \sqrt{a_1 a_3 - (a_1 + b)^2/4} 
\end{align}
and the coefficient $C$ is determined by Eq. \eqref{eq:eq1}.

The thermal effective potential $V_T$ has two parts. Firstly, the usual thermal contributions to the effective potential are given by
\begin{align}
V_{\rm th} = \sum_i \left(\frac{n_{\rm B_i}}{2\pi^2}T^4 J_B \left[\frac{m_{\rm B_i}}{T}\right] \right),
\end{align}
where $T$ is the temperature, and $n_{B_i}$ denotes the degrees of freedom (dof) of the bosonic particles. In general, there exists a fermionic contribution too, but in our model only bosonic contributions exist.
In the above expression, the function $J_B$ is defined as follows:
\begin{align}
&J_B(x) =\int^\infty_0 dz z^2 \log\left[1-e^{-\sqrt{z^2+x^2}}\right] \label{eq:J_B}.
\end{align}
While calculating the thermal potential, we also include a contribution from the daisy diagrams, which constitute the second term in $V_T$. Inclusion of such diagrams improves the perturbative expansion during the phase transition as discussed in earlier works \cite{Fendley:1987ef,Parwani:1991gq,Arnold:1992rz}. There exist two prescriptions to find the daisy improved effective potential by inserting thermal masses into the zero-temperature field dependent masses. In one of these resummation prescriptions, known as the Parwani method \cite{Parwani:1991gq}, thermal corrected field dependent masses are used. In the other prescription, known as the Arnold-Espinosa method \cite{Arnold:1992rz}, the effect of the daisy diagram is included only for Matsubara zero-modes inside $J_B$ function defined above. As we ignore the dark scalar doublet coupling to the SM Higgs, we calculate the field dependent and thermal masses as well as the daisy diagram contributions for $SU(2)_D$ vector bosons only.

As there are two scales of evolution namely, the field $\phi$ itself and temperature $T$ of the universe, we consider the renormalisation scale parameter $u$ instead of $t$ as 
\begin{align}
    u &\equiv \log(\Lambda/M) \quad {\rm where} \quad \Lambda \equiv {\rm max} (\phi,T),
\end{align}
where $\Lambda$ represents the typical scale of the theory. Now, the one-loop level effective potential is given as:
\begin{align}
    V_{\rm tot}(\phi,T) &= V_0(\phi,u) + V_T(\phi,T),   
\end{align}
where 
\begin{align}
    V_T(\phi,T) &= V_{\rm th} + V_{\rm daisy}(\phi,T), \\
    V_{\rm daisy}(\phi,T) &= -\sum_i \frac{g_i T}{12\pi}\left[ m^3_i(\phi,T) - m^3_i(\phi) \right], \nonumber
\end{align}
wherein, $V_{\rm th}$ is the thermal correction and $V_{\rm daisy}$ is the daisy subtraction \cite{Fendley:1987ef,Parwani:1991gq,Arnold:1992rz}. Denoting $m^2_i(\phi,T)=m^2_i(\phi) + \Pi_i(T)$, the relevant thermal masses can be written as \cite{Cline:2008hr}
\begin{align}
    \Pi_{Z_D} = \frac{5}{6} g^2_D T^2, \; \Pi_{\phi, G} = \left( \frac{\lambda_3}{2}+\frac{3}{16} g^2_D \right)T^2.
\end{align}
The parameter $g_i=1, 3, 3$ for $\phi, Z_D, G$ respectively. \\

\begin{figure}
\includegraphics[width=0.45\textwidth]{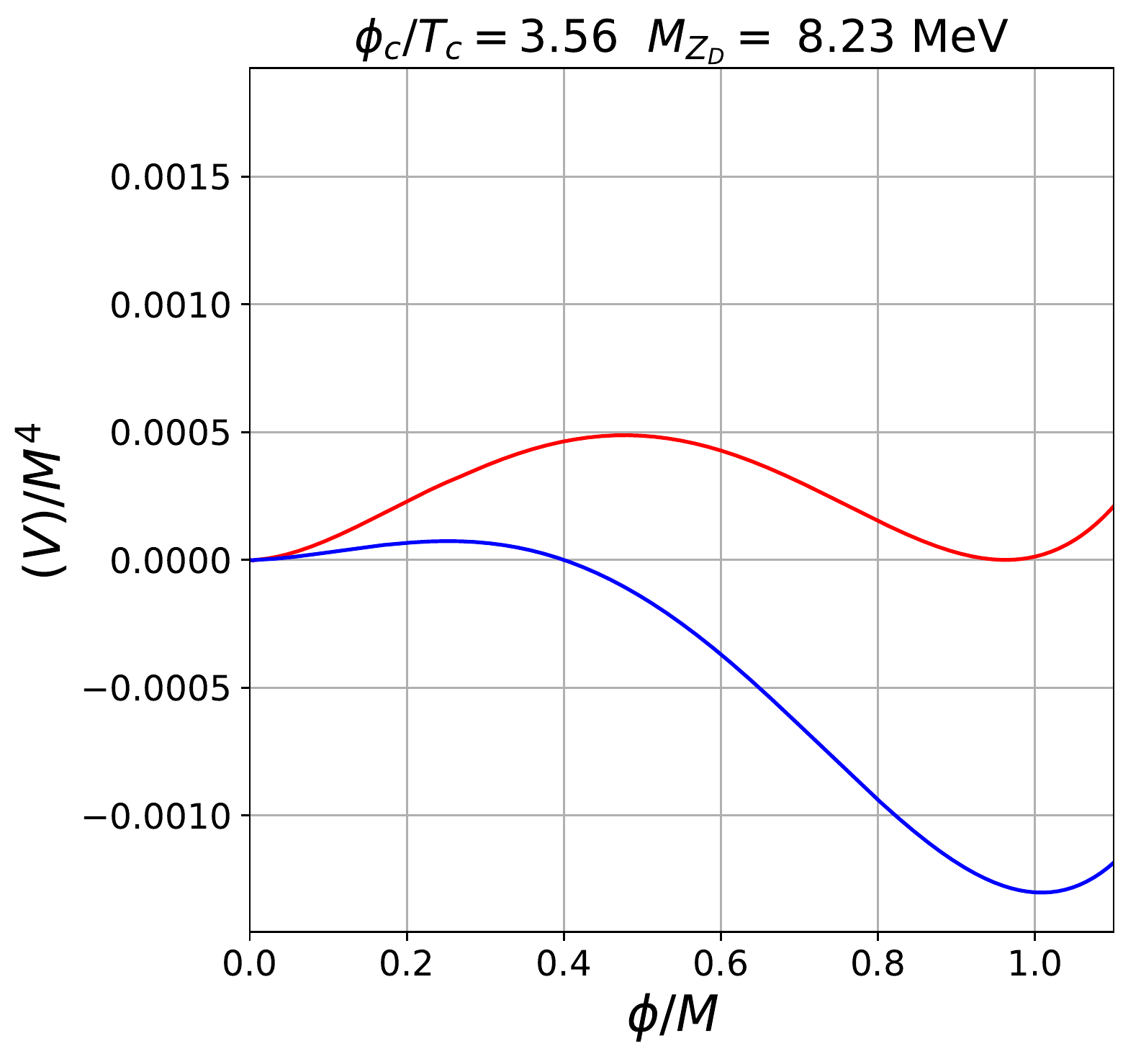}
\caption{Shape of the potential at critical temperature (red curve) and nucleation temperature (blue curve) for chosen benchmark $\alpha_*=0.45, T_*=1.8$ MeV, $g_D=1.37, M_{Z_D}=8.23$ MeV.}
\label{fig0}
\end{figure}

\begin{figure*}
\includegraphics[width=0.45\textwidth]{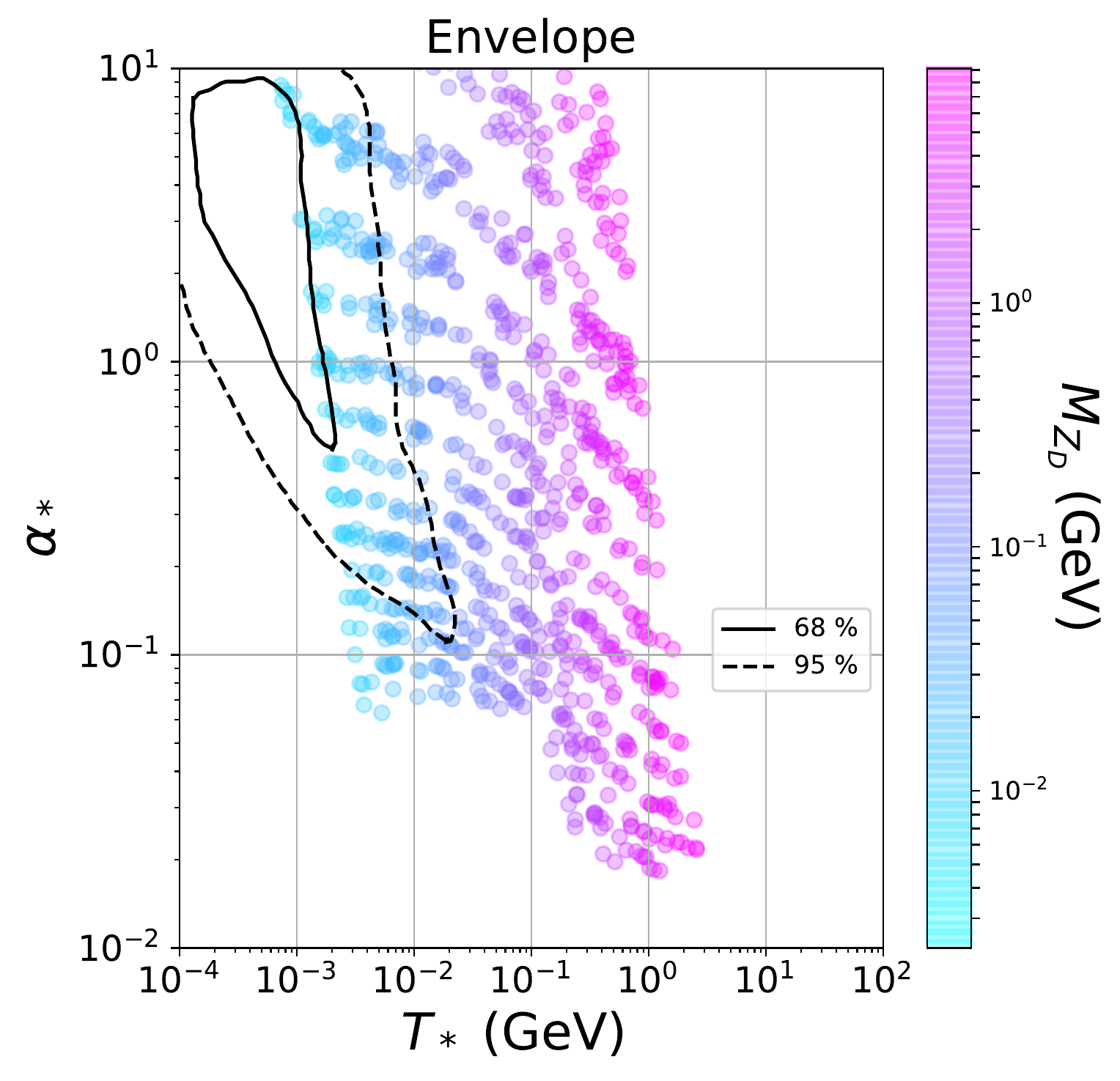}
\includegraphics[width=0.45\textwidth]{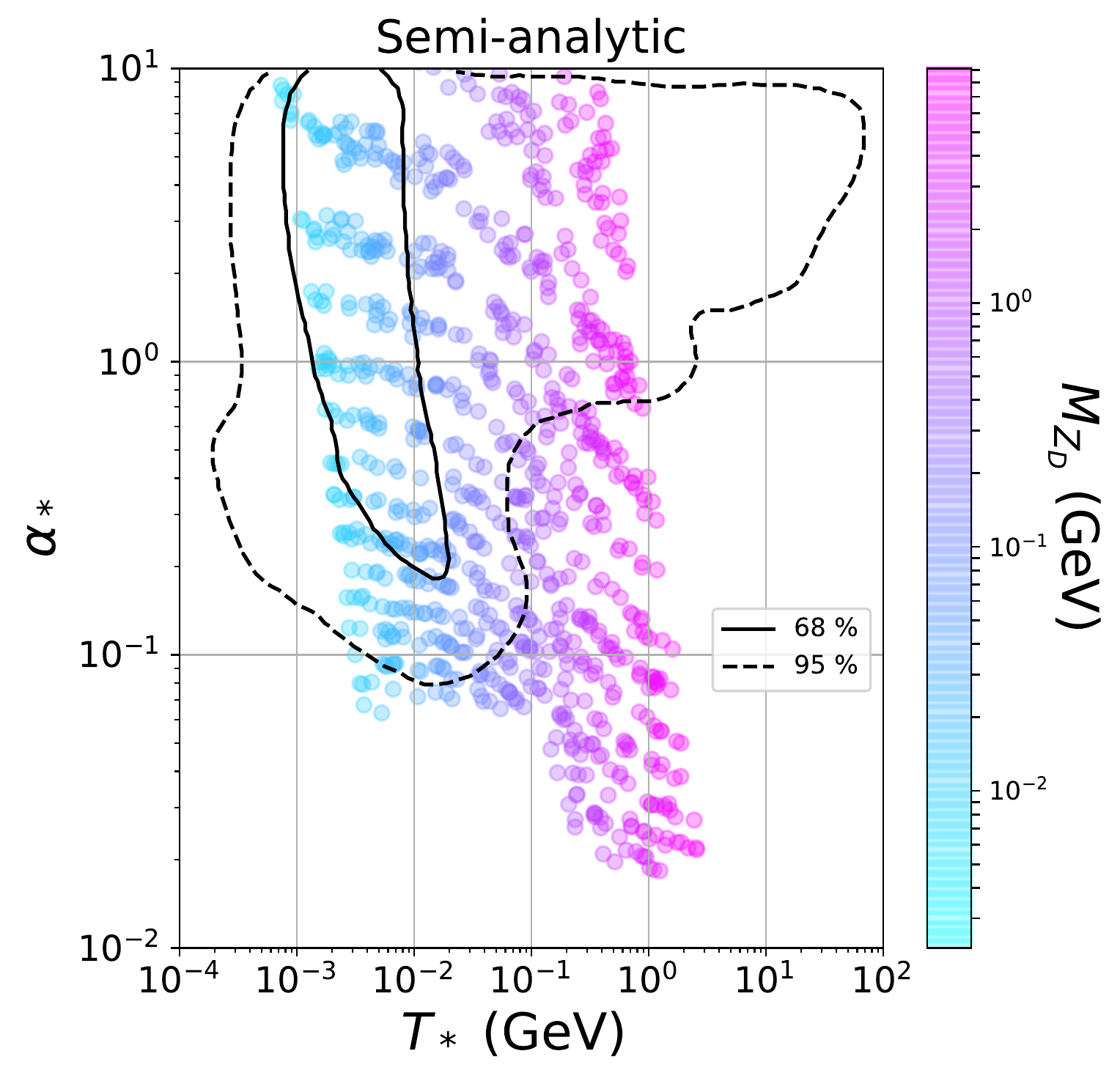}
\includegraphics[width=0.45\textwidth]{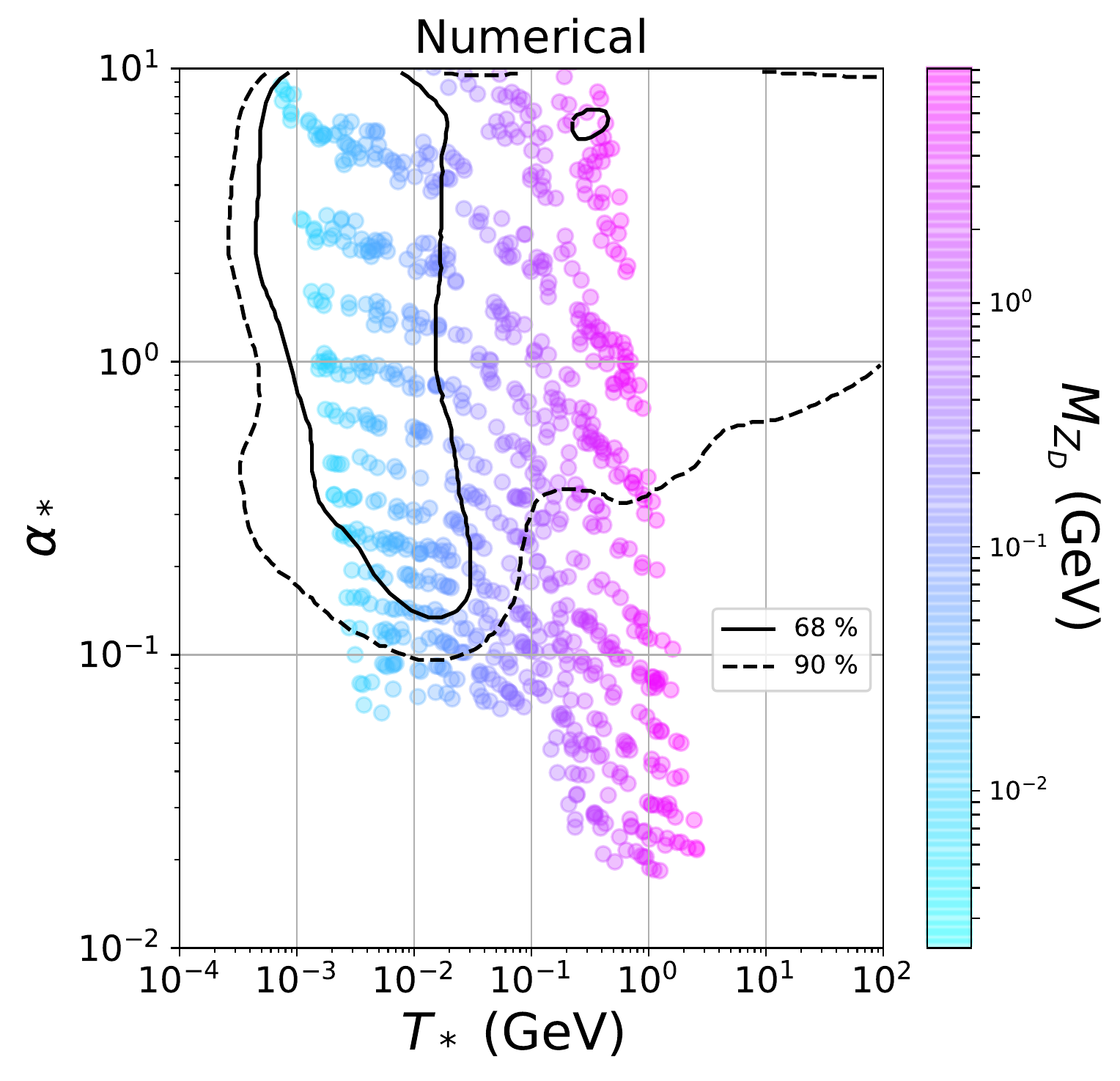}
\caption{The key parameters identifying the $SU(2)_D$ phase transition in the $\alpha_*-T_*$ plane where the $SU(2)_D$ gauge coupling $g_D$ is varied in a range corresponding to $\alpha_D \in 0.01-0.2$ while $M_{Z_D}$ is shown in colour bar. The solid and dashed contours correspond to $68\%$ and $90\%$ confidence levels respectively, according to the NANOGrav analysis \cite{Arzoumanian:2021teu} considering envelope approximation (top left panel), semi-analytic approximation (top right panel), and numerical results (bottom panel).} 
\label{fig1}
\end{figure*}

\begin{figure}
\includegraphics[width=0.45\textwidth]{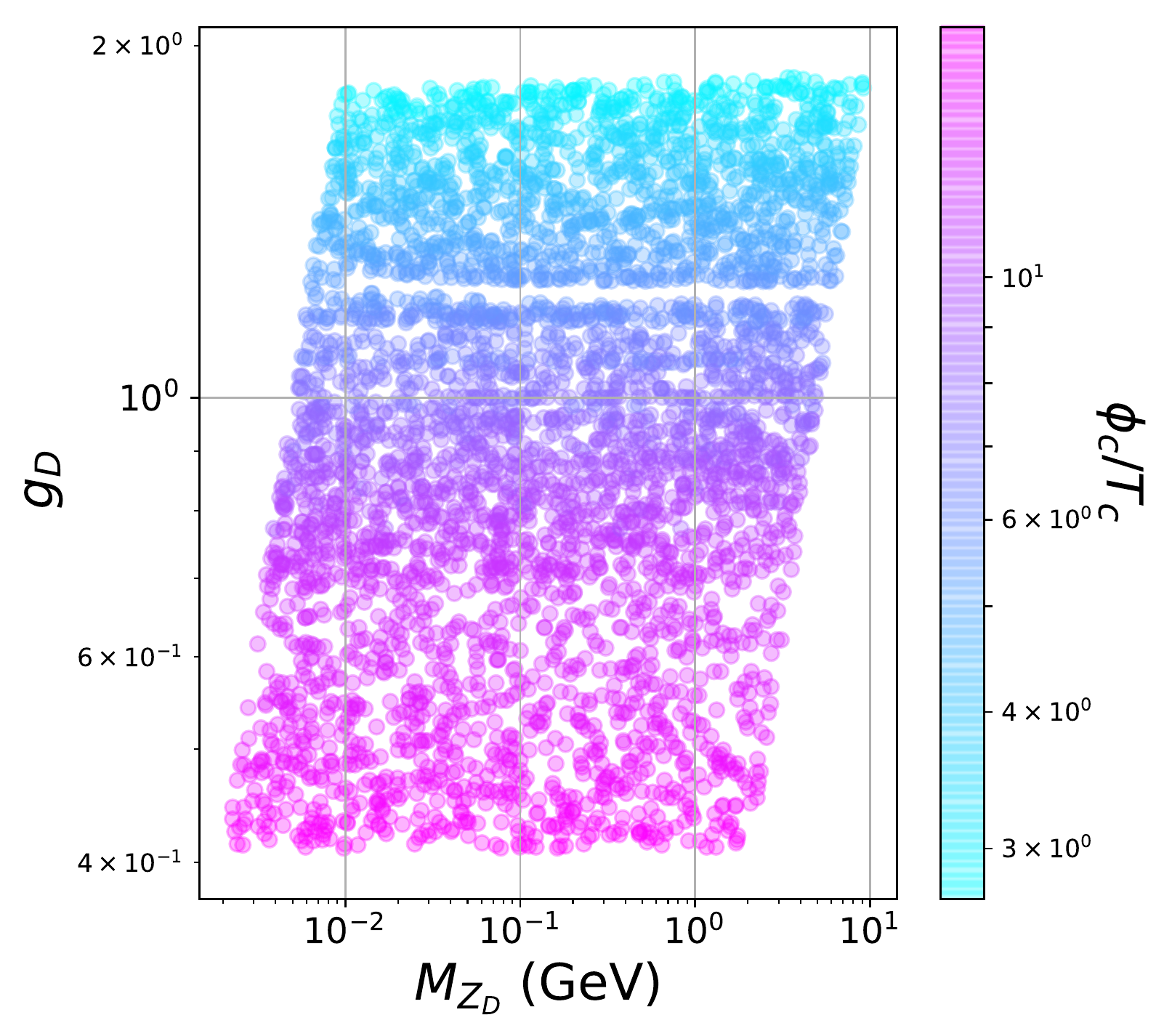}
\caption{The $SU(2)_D$ parameter space in the plane of $g_D$ and $M_{Z_D}$ along with the strength of the FOPT $\phi_c/T_c \equiv \phi(T_c)/T_c$ represented in colour bar.}
\label{fig2}
\end{figure}

\begin{figure}
\includegraphics[width=0.45\textwidth]{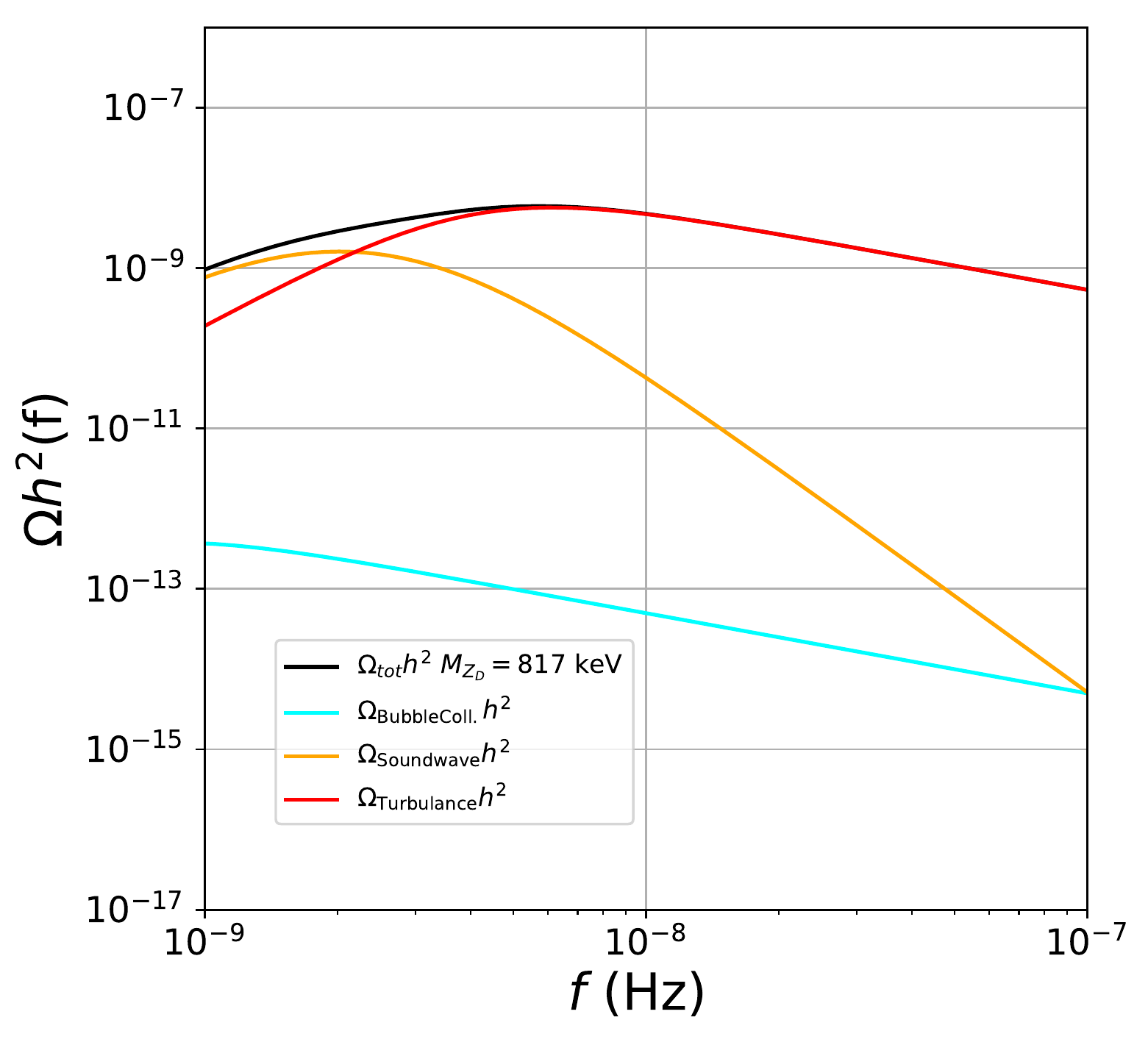}
\caption{GW spectrum $\Omega h^2(f)$ in terms of frequency $f$ for a strong FOPT with benchmark parameters $\alpha_*=0.36, T_*=190$ keV, $g_D=1.37, M_{Z_D}=817$ keV.
The red, orange, cyan and  black curves correspond to
the individual contribution  from  turbulence of the plasma, sound wave of the plasma, bubble collisions, and the total contribution respectively.}
\label{fig3}
\end{figure}

\noindent
{\bf First order phase transition}:
As discussed before, we study the possibility of $SU(2)_D$ phase transition to be of first order. Such transitions proceed via tunnelling with the corresponding spherical symmetric field configurations known as bubbles being nucleated followed by expansion and coalescence. Recent reviews of cosmological phase transitions can be found in \cite{Mazumdar:2018dfl,Hindmarsh:2020hop}. The tunnelling rate per unit time per unit volume is given as \cite{Linde:1981zj}
\begin{align}
\Gamma (T) = \mathcal{A}(T) e^{-S_3(T)/T},
\end{align}
where $\mathcal{A}(T)\sim T^4(S_3/(2\pi T))^{3/2}$ and $S_3(T)$ are determined by the dimensional analysis and by the classical configurations, called bounce, respectively. At finite temperature, the $O(3)$ symmetric bounce solution \cite{Linde:1980tt} corresponds to the solution of the differential equation given as follows,
\begin{align}
    \frac{d^2 \phi}{dr^2}+\frac{2}{r}\frac{d\phi}{dr} = \frac{\partial V_{\rm tot}}{\partial \phi}\label{eq:bounce diff}.
\end{align}
This equation can be solved by using the boundary conditions given by
\begin{align}
\phi(r\to \infty)= \phi_{\rm false},~~~\left.\frac{d\phi}{dr}\right|_{r=0} =0,\label{eq:boundary condition}
\end{align}
where $\phi_{\rm false}$ denotes the position of the false vacuum. Using $\phi$ governed by the above equation and boundary conditions, the bounce action can be written as
\begin{align}
    S_3 =\int_0^{\infty} dr 4\pi r^2 \left[\frac{1}{2}\left(\frac{d\phi}{dr}\right)^2 +V_{\rm tot}(\phi,T)\right].
    \label{s3eq}
\end{align}
The temperature corresponding to bubble nucleation is known as the nucleation temperature $T_*$ which is calculated by comparing the tunnelling rate to the Hubble expansion rate as
\begin{align}
    \Gamma (T_*) = {\bf H}^4(T_*).
\end{align}
Considering a radiation dominated universe, the Hubble parameter is given by ${\bf H}(T)\simeq 1.66\sqrt{g_*}T^2/M_{\rm Pl}$ with $g_*$ being the dof of the radiation component. 
The critical temperature $T_c$ corresponds to the temperature where the two minima of the potential are degenerate. Now, if we have $\phi(T_c)/T_c > 1$ where $\phi(T_c)$ is the dark scalar VEV at the critical temperature $T=T_c$, the corresponding phase transition is conventionally called a \textit{strong} first order phase transition (SFOPT). Sometimes, this criteria is also referred to as $\phi(T_*) / T_*>1$, where $\phi(T_*)$ is the dark scalar VEV at the nucleation temperature, $T=T_*$. We consider $\phi(T_c)/T_c > 1$ in our work which usually guarantees the validity of the latter.

The free energy difference between the true and the false vacuum is given by
\begin{equation}
    \Delta V_{\rm tot} \equiv V_{\rm tot}(\phi_{\rm false},T)- V_{\rm tot}(\phi_{\rm true},T).
\end{equation}
As a result of the bubble nucleation, the amount of vacuum energy released by the phase transition, in the units of radiation energy density of the universe, $\rho_{\rm rad}= g_*\pi^2 T^4/30 $, is given by
\begin{align}
    \alpha_* =\frac{\epsilon_*}{\rho_{\rm rad}},
    \label{alphastar}
\end{align}
with
\begin{align}
    \epsilon_* = \left[\Delta V_{\rm tot} - \frac{T}{4} \frac{\partial \Delta V_{\rm tot}}{\partial T}\right]_{T=T_*},
\end{align}
which is also related to the change in the trace of the energy-momentum tensor across the bubble wall \cite{Caprini:2019egz,Borah:2020wut}.

In Fig.~\ref{fig0}, we show the shape of the potential in terms of $\phi/M$ at critical $(T_c)$ and nucleation $(T_*)$ temperatures. For illustrative purpose, we choose the relevant benchmark values as $\alpha_*=0.45, T_*=1.8$ MeV, $g_D=1.37$, and $M_{Z_D}=8.23$ MeV. The red coloured curve corresponds to the potential at $T=T_c$, while the one in blue colour corresponds to $T=T_*$. Clearly, $\phi=0$ becomes a false vacuum below the critical temperature $T_c$. Also, from the shape of the potential at $T_c$, it can be clearly seen that there exists a potential barrier between the two vacua, an indication of a SFOPT. This eventually triggers bubble production and subsequent production of GW.

The SFOPT discussed above gets completed via the percolation of
the growing bubbles. In order to determine the epoch of completion of the phase transition, one needs to estimate
the percolation temperature $T_p$ at which significant volume of the universe is converted from the symmetric phase (false vacuum) to the broken phase (true vacuum). Adopting the prescription given in \cite{Ellis:2018mja, Ellis:2020nnr},
the percolation temperature $T_p$ is obtained from the probability of finding a point still in the false vacuum
given by
%
\begin{align}
    \mathcal{P}(T) = e^{-\mathcal{I}(T)}, \nonumber
\end{align}  
where
\begin{align}
    \mathcal{I}(T) = \frac{4\pi}{3}\int^{T_c}_T \frac{dT'}{T'^4}\frac{\Gamma(T')}{{\bf H}(T')}\left(\int^{T'}_T \frac{d\tilde{T}}{{\bf H}(\tilde{T})}\right)^3.
\end{align}
The percolation temperature is then calculated by using  $\mathcal{I}(T_p) = 0.34$ \cite{Ellis:2018mja} (implying that at least $34\%$ of the comoving volume is occupied by the true vacuum). It is important to ensure that the physical volume of the false vacuum gets decreased around percolation for successful completion of the phase transition. This requirement gives rise to the following condition \cite{Ellis:2018mja, Ellis:2020nnr}
\begin{align}
    \frac{1}{\mathcal{V}_{\rm false}}\frac{d\mathcal{V}_{\rm false}}{dx}&={\bf H}(T)\left(3+T\frac{d\mathcal{I}(T)}{dT}\right)<0. \quad ; x \equiv {\rm time} \label{eq:cond}
\end{align}
ensuring which, at the percolation temperature $T_p$, can confirm the successful completion of the phase transition. For some chosen benchmark values, including the one shown in Fig. \ref{fig0}, we have confirmed the validity of the above condition at the percolation temperature $T_p$, as summarised in table \ref{tab:GWBP}. \\

\noindent
{\bf Gravitational wave}: A SFOPT can lead to the generation of stochastic GW background primarily due to three mechanisms namely, the bubble collisions~\cite{Turner:1990rc,Kosowsky:1991ua,Kosowsky:1992rz,Kosowsky:1992vn,Turner:1992tz}, the sound wave of the plasma~\cite{Hindmarsh:2013xza,Giblin:2014qia,Hindmarsh:2015qta,Hindmarsh:2017gnf} and the turbulence of the plasma~\cite{Kamionkowski:1993fg,Kosowsky:2001xp,Caprini:2006jb,Gogoberidze:2007an,Caprini:2009yp,Niksa:2018ofa}.

The amplitudes of such GW signal depend upon the the amount of vacuum energy released by the phase transition in comparison to the radiation energy density of the universe, $\rho_{\rm rad}= g_*\pi^2 T^4/30 $, given by $\alpha_*$ defined in Eq. \eqref{alphastar}. The amplitude of GW also depends upon the duration of the FOPT, denoted by the parameter $\beta$, defined as \cite{Caprini:2015zlo}
\begin{align}
\frac{\beta}{{\bf H}(T)} \simeq T\frac{d}{dT} \left(\frac{S_3}{T} \right).
\end{align}
Here, $\alpha_*$ and $\beta/{\bf H}(T)$ are evaluated at the nucleation temperature $T=T_*$. While $S_3$ can be evaluated using Eq. \eqref{s3eq}, the effective potential at sufficiently low temperatures i.e., $T \ll M$, can be safely approximated as 
\begin{align}
    V_{\rm tot} &\simeq \frac{g^2_{D}(t^{\prime})}{2}T^2\phi^2 + \frac{\lambda_{\rm eff}(t^{\prime})}{4}\phi^4  \label{eq:Veffapp},\\
  {\rm with}~~  \lambda_{\rm eff}(t') &= \frac{4\pi \alpha_\lambda(t')}{\left(1-\frac{b}{2\pi}\alpha_{D}(0)t'\right)^{a_2/b}}, \nonumber \\
    t^{\prime} &= \ln(T/M).\nonumber 
\end{align}
With such approximation, the action can be written as \cite{Linde:1981zj, Jinno:2016knw}
\begin{align}
    S &= \frac{S_3}{T} - 4\ln(T/M), \nonumber \\
    \frac{S_3}{T} &\simeq -9.45 \times \frac{g_{D}(t')}{\lambda_{\rm eff}(t')}.
    \label{eq:s3app}
\end{align}
While estimating the GW amplitude we have used the expressions given by Eq. \eqref{eq:Veffapp} and Eq. \eqref{eq:s3app} to $\alpha$, $\beta$ and the percolation temperature $T_p$. We have cross-checked the validity of the approximated analytical expression for $S_3/T$ mentioned above for the benchmark points mentioned in table \ref{tab:GWBP} using numerical packages {\tt SimpleBounce}\cite{Sato:2019wpo} and {\tt BubbleProfiler} \cite{Athron:2019nbd} and found that the results from numerical analysis and those from the approximated expression differ by up to 10 $\%$.

\begin{figure}
\includegraphics[width=0.45\textwidth]{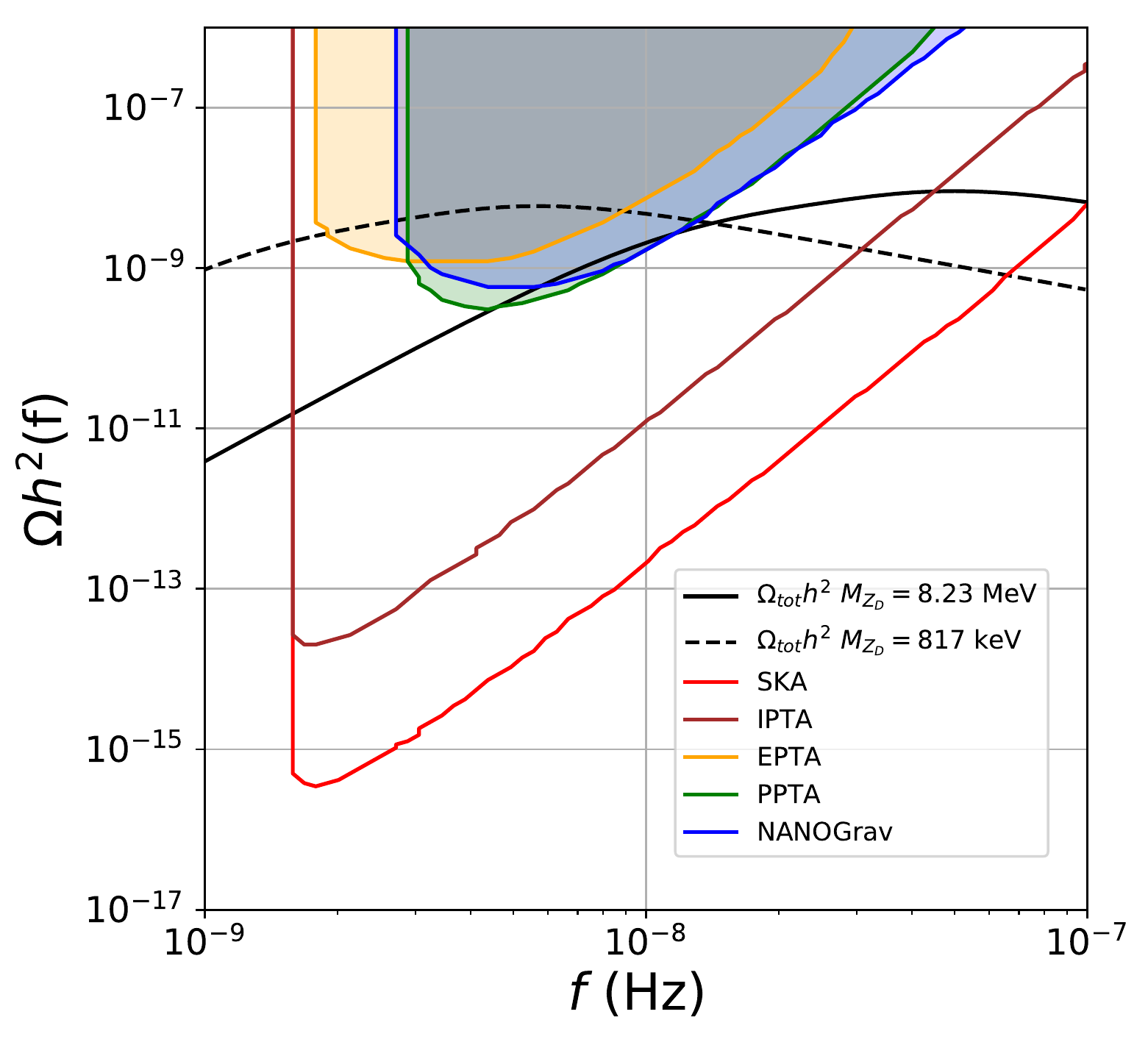}
\caption{GW spectrum $\Omega h^2(f)$ in terms of frequency $f$ for a strong FOPT. The black solid and black dashed curves correspond to the benchmark parameters shown in table \ref{tab:GWBP}. The other contours and shaded regions correspond to sensitivities of several different experiments.}
\label{fig4}
\end{figure}

The NANOGrav collaboration, in their analysis \cite{Arzoumanian:2021teu}, has estimated the required FOPT parameters using thin shell approximation for bubble walls (envelope approximation) \cite{Jinno:2016vai}, semi-analytic approximation \cite{Lewicki:2020azd} as well as full lattice results. Here, we present the predictions of our model (coloured points) against the backdrop of their estimates in Fig. \ref{fig1}.
 In the analysis, the gauge coupling $g_D$ is varied in a range corresponding to $0.01\lesssim \alpha_D \lesssim 0.2$, while the gauge boson masses are shown in the colour bar. The solid ( dashed) contour corresponds to the allowed region at $68 (95)\%$ confidence level obtained in \cite{Arzoumanian:2021teu} by using envelope approximation (top left panel), semi-analytic approximation (top right panel), and numerical results (bottom panel). Clearly, predictions based on light gauge boson with masses in sub-GeV regime are in better agreement with the NANOGrav findings. In order to see the strength of the FOPT in terms of $SU(2)_D$ parameters, we plot $g_D$ with respect to $M_{Z_D}$ along with  $\phi_c/T_c \equiv \phi(T_c)/T_c$ whose values are varied as shown in colour bar in Fig. \ref{fig2}. As can be seen from the colour bar, the strength of the phase transition $\phi_c/T_c$ increases slightly by a numerical factor when $g_D$ is decreased by a similar factor keeping $M_{Z_D}$ fixed.

As mentioned above, there are three sources of GW production during a FOPT: bubble collisions, sound wave of the plasma, and turbulence of the plasma \cite{Jinno:2016vai, Caprini:2009yp, Hindmarsh:2017gnf, Binetruy:2012ze,Hindmarsh:2015qta,Caprini:2015zlo}. 
Therefore,  following \cite{Arzoumanian:2021teu}, the resultant GW power spectrum can be written as
\begin{align}
    \Omega_{\rm GW}(f) &= \Omega_\phi(f) + \Omega_{\rm sw}(f) + \Omega_{\rm turb}(f).
\end{align}
In general, each contribution can be characterised by its own peak frequency and each GW spectrum can be parametrised, following \cite{Arzoumanian:2021teu}, as
\begin{align}
    h^2\Omega(f) &= \mathcal{R}\Delta(v_w)\left(\frac{\kappa \alpha_*}{1+\alpha_*}\right)^p\left(\frac{{\bf H_*}}{\beta}\right)^*\mathcal{S}(f/f^0_*),
\end{align}
where the pre-factor $\mathcal{R}\simeq 7.69\times 10^{-5}g^{-1/3}_*$ takes in account the red-shift of the GW energy density, $\mathcal{S}(f/f^0_*)$ parametrises the shape of the spectrum and $\Delta(v_w)$ is the normalization factor which depends on the bubble wall velocity $v_w$. The Hubble parameter at the nucleation temperature $T=T_*$ is denoted by ${\bf H_*}$. Finally the peak frequency today, $f^0_*$, is related to the value of the peak frequency at the time of emission, $f_*$, as follows
\begin{align}
    f^0_* &\simeq 1.13\times 10^{-10}{\rm Hz} \left(\frac{f_*}{\beta}\right)\left(\frac{\beta}{{\bf H_*}}\right)\left(\frac{T_*}{ {\rm MeV}}\right)\left(\frac{g_*}{10}\right)^{1/6},
\end{align}
where $g_*$ denotes the number of relativistic dof at the time of the FOPT. The values of the peak frequency at the time of emission, the normalisation factor, the spectral shape, and the exponents $p$ and $q$ are given in Table I of \cite{Arzoumanian:2021teu}. The efficiency factors namely, $\kappa_\phi$ is discussed in \cite{Jinno:2016knw, Ellis:2020nnr} and $\kappa_{\rm sw}$ is taken from \cite{Espinosa:2010hh,Borah:2020wut}. On the other hand, the remaining efficiency factor $\kappa_{\rm turb}$ is taken to be approximately $0.1 \times \kappa_{\rm sw}$ \cite{Arzoumanian:2021teu}. The details of bubble wall velocities and their values can be found in \cite{Steinhardt:1981ct, Huber:2013kj,Leitao:2014pda,Dorsch:2018pat,Cline:2020jre, Azatov:2020ufh}.

\begin{table}[h!]
    \centering
    \begin{tabular}{|c|c|c|c|c|c|}
         \hline 
        $\alpha_*$ &  $(\beta/{\bf H}_*)$ & $T_*$ & $v_w$ & $T_p$ & $ \frac{1}{\mathcal{V}_{\rm false}} \frac{d\mathcal{V}_{\rm false}}{dx}$\\
         \hline
        0.45 & 143 & 1.8 MeV & 0.887 & 8.23 MeV & $-1.3\times 10^{-19}$ GeV\\
        \hline
        0.36 & 151 & 190 keV & 0.872 & 817 keV & $-6.88\times 10^{-18}$ GeV\\
         \hline
    \end{tabular}
    \caption{Numerical values of benchmark parameters used in the estimation of GW spectrum.}
    \label{tab:GWBP}
\end{table}

\begin{figure*}
\includegraphics[width=0.45\textwidth]{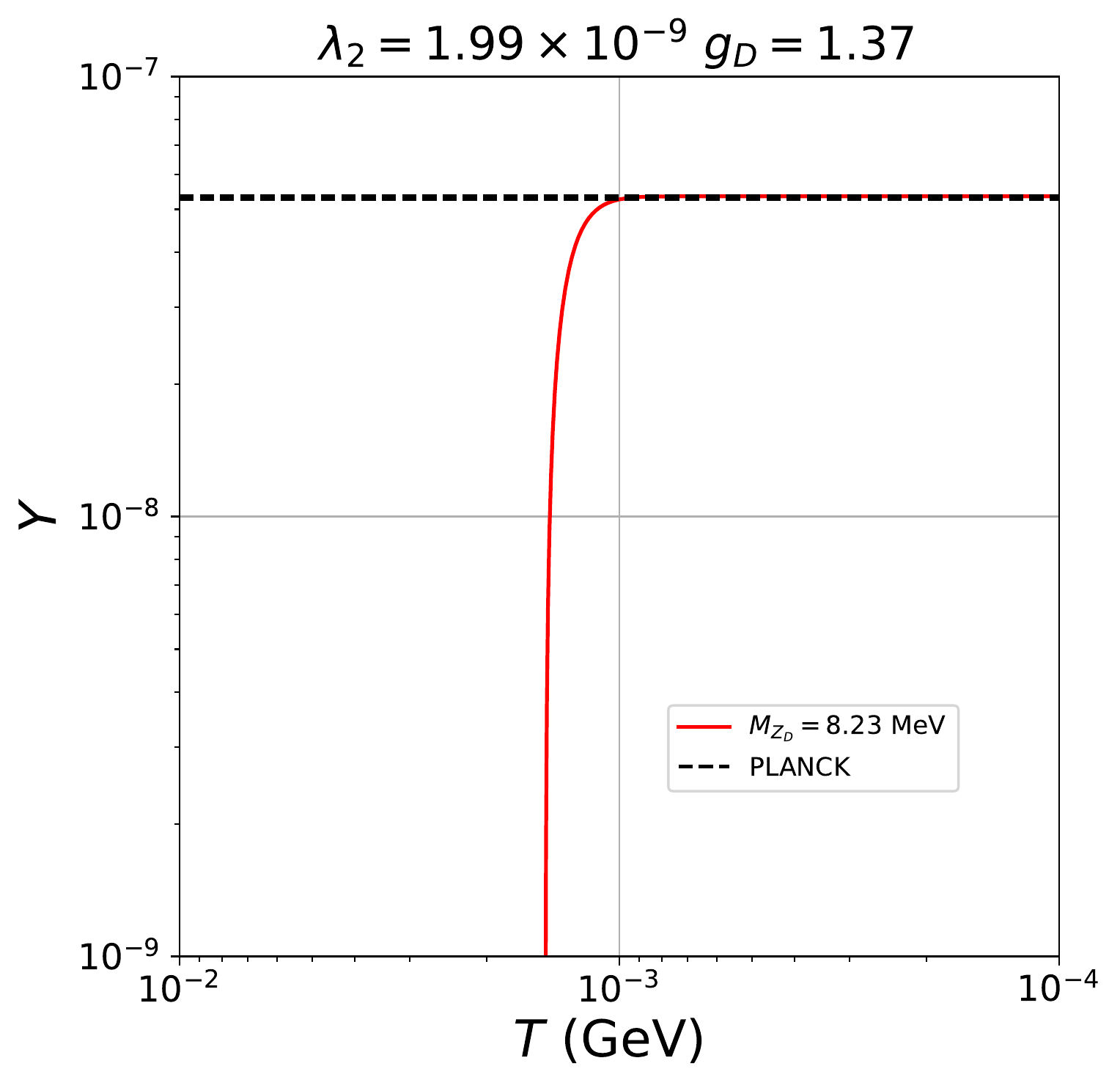}
\includegraphics[width=0.45\textwidth]{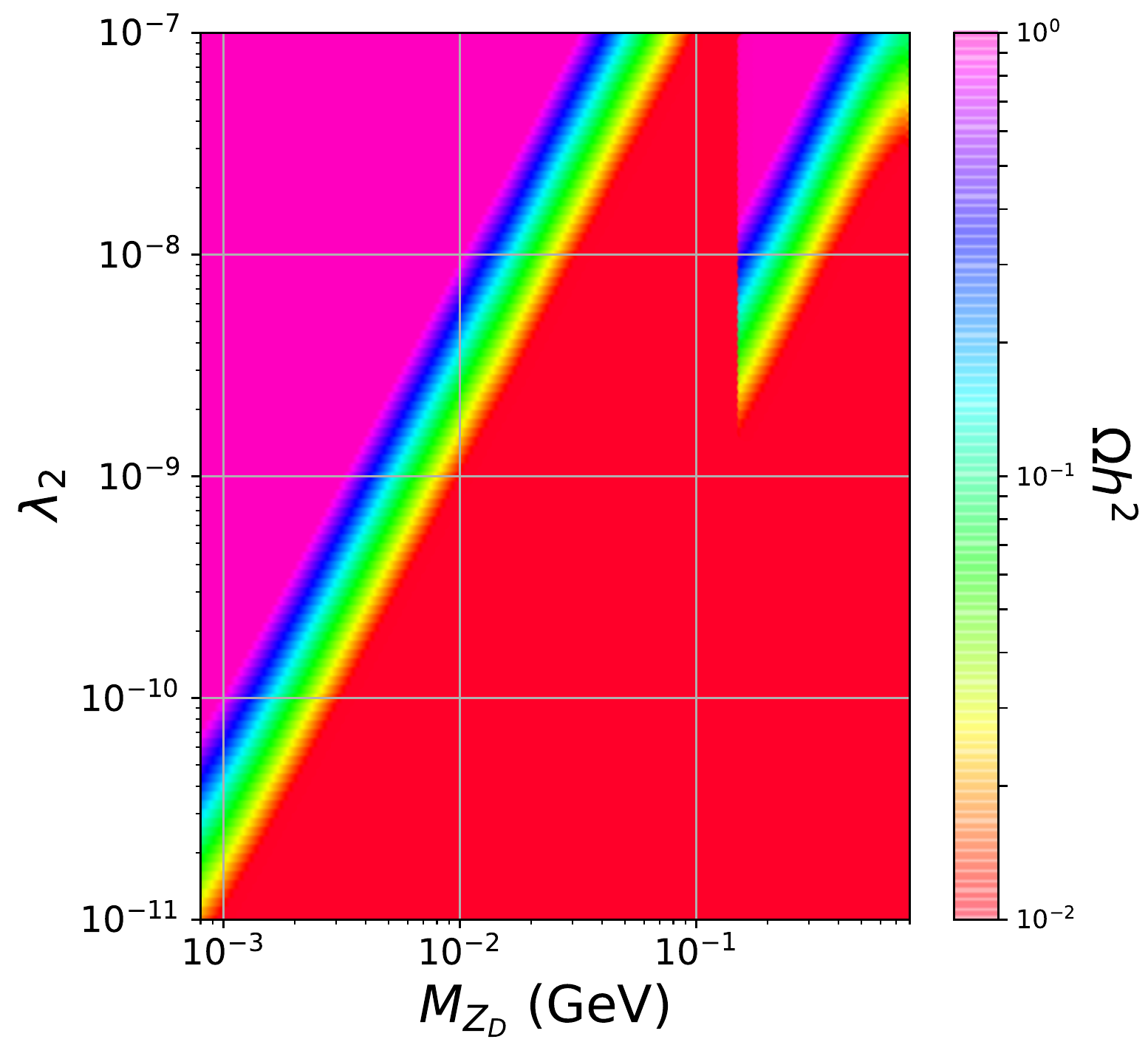}
\caption{Left panel: Comoving DM number density $Y$ as a function of temperature $T$ for DM mass $M_{Z_{\rm D}}=8.23$ MeV and $\lambda_2 = 1.99 \times 10^{-9}$. Right panel: The regions of $M_{Z_{\rm D}}-\lambda_2$ parameter space giving rise to DM relic (shown in colour bar) for $g_D=1.37$.} 
\label{fig5}
\end{figure*}

Using the above-mentioned prescription for estimating GW spectrum from a strong FOPT and by choosing a benchmark values of model
as well as FOPT parameters shown in table \ref{tab:GWBP} consistent with NANOGrav data at $95\%$ CL, we calculate 
the individual contributions to GW energy density spectrum $\Omega h^2(f)$ from bubble collisions, sound wave of the plasma, and turbulence of the plasma as well as the total contribution to $\Omega h^2(f)$.
In Fig. \ref{fig3}, 
the red, orange, cyan and  black curves correspond to
the individual contribution  from  turbulence of the plasma, sound wave of the plasma, bubble collisions, and the total contribution to $\Omega h^2(f)$, respectively.
Due to the small value of FOPT strength parameter $\alpha_*$, as anticipated 
from earlier studies \cite{Bodeker:2017cim,Ellis:2019oqb},
the contribution from bubble collision remains suppressed compared to the other two contributions as can be seen in Fig. \ref{fig3}. In Fig. \ref{fig4}, we show the GW spectrum for two benchmark points shown in table \ref{tab:GWBP}. The sensitivities of different PTA based experiments \cite{Schmitz:2020syl} like NANOGrav \cite{McLaughlin:2013ira, Arzoumanian:2021teu, Arzoumanian:2020vkk} , SKA \cite{Weltman:2018zrl}, IPTA \cite{Hobbs_2010}, PPTA \cite{Manchester_2013}, EPTA \cite{Kramer_2013} are also shown in Fig. \ref{fig4}. Clearly, heavier $Z_D$ leads to shift in the peak frequency towards larger values, as expected. While the benchmark point with heavier $Z_D$ remains barely within NANOGrav reach, the future experiments are sensitive to both the benchmark points over a much wider range of frequencies. \\

\noindent
{\bf Dark matter}: Origin of particle DM has been a longstanding puzzle. Ordinary or visible matter contributes only one-fifth $(\sim 20\%)$ to the total matter content of the present universe. The rest of the matter remain in the form of a mysterious, non-luminous, non-baryonic component, often referred to as DM. While there have been astrophysical evidences for such non-baryonic matter for several decades \cite{Zwicky:1933gu, Rubin:1970zza, Clowe:2006eq}, precision measurements of the cosmic microwave background (CMB) anisotropies at cosmology experiments like WMAP, Planck have confirmed its existence in a convincing way. The present abundance of DM is often quantified in terms of a dimensionless quantity as \cite{Aghanim:2018eyx}:
\begin{equation}
\Omega_{\text{DM}} h^2 = 0.120\pm 0.001
\label{dm_relic}
\end{equation}
at 68\% confidence level (CL). Here $\Omega_{\rm DM}=\rho_{\rm DM}/\rho_{\rm critical}$ is the density parameter of DM and $h = \text{Hubble Parameter}/(100 \;\text{km} ~\text{s}^{-1} 
\text{Mpc}^{-1})$ is a dimensionless parameter of order unity. $\rho_{\rm critical}=3H^2/(8\pi G)$ is the critical density while $H$ is the Hubble parameter. Among different particle DM proposals in the literature,  the weakly interacting massive particle (WIMP) paradigm is the most appealing one. In such a scenario, a stable or sufficiently long-lived DM particle having mass and interaction strength typically around the electroweak corner can get produced thermally from the SM bath in early universe, followed by freeze-out from the bath, leaving a relic similar to the observed DM abundance \cite{Kolb:1990vq}. Apart from this remarkable coincidence referred to as the {\it WIMP Miracle}, such DM can be probed at direct detection experiments by virtue of their sizeable interactions with SM particles like quarks \cite{Arcadi:2017kky}. However, due to absence of any such signals, alternatives to WIMP have also been discussed in recent times. One such appealing alternative is the non-thermal origin of DM, known as the feebly interacting (or freeze-in) massive particle (FIMP) DM \cite{Hall:2009bx, Bernal:2017kxu}. In such a scenario, DM has so feeble interactions with the SM particles that it never reaches thermal equilibrium, but gets produced non-thermally due to decay or scattering of SM bath particles. 

As mentioned before, here we consider the $SU(2)_D$ vector bosons as DM candidates. This has been explored in several earlier works \cite{Diaz-Cruz:2010czr, Fraser:2014yga, Bhattacharya:2011tr, Barman:2017yzr, Barman:2018esi, Baldes:2018emh, Barman:2019lvm, Abe:2020mph, Nomura:2020zlm, Chowdhury:2021tnm}. While most of these works considered thermal vector boson DM, the non-thermal or FIMP possibility was discussed in \cite{Barman:2019lvm}. The scenario in our present model is much more simpler as SM-DM interactions occur only via the Higgs portal. As we had assumed tiny Higgs portal coupling $\lambda_2$ between dark scalar $\Phi$ and SM Higgs doublet $H$ while discussing the FOPT details, it naturally provides the freeze-in portal. Additionally, constraints from CMB measurements disfavour light sub-GeV thermal DM production in the early universe through s-channel annihilations into SM fermions \cite{Aghanim:2018eyx}. Therefore, we stick to the non-thermal DM scenario here.

In general, the Boltzmann equation for comoving DM density in FIMP scenarios can be written as
\begin{equation}
\dfrac{dY}{dx^{\prime}}=\dfrac{\langle\sigma v\rangle\,s}{{\bf H}\,x}(Y^{\rm eq}_{\rm SM})^2
\end{equation}
with $x^{\prime}=M_{\text{DM}}/T$ and $s = \frac{2\pi^2}{45}g_{*s} T^3$ being the entropy density. In the above equation, we consider the freeze-in DM production from SM bath via scatterings of the type ${\rm SM \; SM} \rightarrow {\rm DM \; DM}$ with thermal averaged cross-section denoted by $\langle \sigma v \rangle$. Such a scattering is mediated by scalars via $H-\Phi$ mixing. In Fig.~\ref{fig5} (left panel), we show the evolution of comoving DM density $Y$ as a function of temperature for a benchmark choice of DM mass $M_{\rm DM}=M_{Z_D}=8.23$ MeV, $\lambda_2 = 1.99 \times 10^{-9}$, and $g_D=1.37$. While dark gauge coupling is of order one, the smallness of $\lambda_2$ leads to a tiny scalar portal mixing required for realising FIMP scenario. It should be noted that the DM abundance rises sharply around a temperature close to its mass. This is because the scalar portal mixing arises dynamically only after the $SU(2)_D$ symmetry breaking at a scale $\sim M_{Z_D}/g_D$. On the right panel of Fig. \ref{fig5}, we show the parameter space in $\lambda_2-M_{Z_D}$ plane with DM relic as shown in the colour bar. The parameter space consistent with correct DM relic abundance shows a linear relation between $\lambda_2$ and $M_{Z_D}$. This can be understood as follows. Since $g_D$ as well as $\Phi$ quartic coupling $\lambda_3$ are fixed, larger $M_{Z_D}$ implies larger VEV of $\Phi$ and hence larger scalar mass mediating SM-DM interactions. To compensate for this larger mediator mass, a larger mixing (and hence a larger $\lambda_2$) is required to generate the correct DM relic. The sharp discontinuity after DM mass crosses muon mass threshold arises as it corresponds to the FOPT occurring at temperatures above the muon mass threshold, allowing muon in the radiation bath to also contribute enhancing the contribution to DM freeze-in. \\

\noindent
{\bf Conclusion}: We have studied a minimal $SU(2)_D$ gauge extension of the SM with the possibility of a strong first order phase transition within the dark sector, specially at a low temperature below the EW scale such that resulting stochastic GWs can be observed at PTA based experiments like NANOGrav. Motivated by the recent results from the NANOGrav 12.5 yr data providing hints of such a cosmological phase transition at sub-EW scale, we constrain our model parameters from the requirement of fitting NANOGrav data. A SFOPT occurring at sub-GeV scale can explain the NANOGrav data very well while also being sensitive to future experiments to be operating in this low frequency regime of GWs. The non-Abelian nature of our dark gauge symmetry also provides a natural vector boson DM candidate which is naturally stable due to the absence of any kinetic mixing with $U(1)_Y$ of the standard model at renormalisable level. Such light vector boson DM can be produced via freeze-in mechanism in the early universe. While such freeze-in mechanism is naturally realised due to tiny scalar portal couplings required to realise dark SFOPT without disturbing the EW vacuum and vice versa, it also helps in avoiding stringent CMB bounds on such light DM, if its production happens thermally. Depending upon the size of scalar portal couplings, we can have more non-trivial FOPT as well DM phenomenology which we leave for future studies.

\noindent
\acknowledgements
DB acknowledges the support from Early Career Research Award from DST-SERB, Government of India (reference number: ECR/2017/001873). AD and SKK are supported in part by the National Research Foundation (NRF) grants NRF-2019R1A2C1088953.


\begin{thebibliography}{90}
\expandafter\ifx\csname natexlab\endcsname\relax\def\natexlab#1{#1}\fi
\expandafter\ifx\csname bibnamefont\endcsname\relax
  \def\bibnamefont#1{#1}\fi
\expandafter\ifx\csname bibfnamefont\endcsname\relax
  \def\bibfnamefont#1{#1}\fi
\expandafter\ifx\csname citenamefont\endcsname\relax
  \def\citenamefont#1{#1}\fi
\expandafter\ifx\csname url\endcsname\relax
  \def\url#1{\texttt{#1}}\fi
\expandafter\ifx\csname urlprefix\endcsname\relax\def\urlprefix{URL }\fi
\providecommand{\bibinfo}[2]{#2}
\providecommand{\eprint}[2][]{\url{#2}}

\bibitem[{\citenamefont{Arzoumanian et~al.}(2021)}]{Arzoumanian:2021teu}
\bibinfo{author}{\bibfnamefont{Z.}~\bibnamefont{Arzoumanian}}
  \bibnamefont{et~al.} (\bibinfo{year}{2021}), \eprint{2104.13930}.

\bibitem[{\citenamefont{Arzoumanian et~al.}(2020)}]{Arzoumanian:2020vkk}
\bibinfo{author}{\bibfnamefont{Z.}~\bibnamefont{Arzoumanian}}
  \bibnamefont{et~al.} (\bibinfo{collaboration}{NANOGrav}),
  \bibinfo{journal}{Astrophys. J. Lett.} \textbf{\bibinfo{volume}{905}},
  \bibinfo{pages}{L34} (\bibinfo{year}{2020}), \eprint{2009.04496}.

\bibitem[{\citenamefont{Blasi et~al.}(2021)\citenamefont{Blasi, Brdar, and
  Schmitz}}]{Blasi:2020mfx}
\bibinfo{author}{\bibfnamefont{S.}~\bibnamefont{Blasi}},
  \bibinfo{author}{\bibfnamefont{V.}~\bibnamefont{Brdar}}, \bibnamefont{and}
  \bibinfo{author}{\bibfnamefont{K.}~\bibnamefont{Schmitz}},
  \bibinfo{journal}{Phys. Rev. Lett.} \textbf{\bibinfo{volume}{126}},
  \bibinfo{pages}{041305} (\bibinfo{year}{2021}), \eprint{2009.06607}.

\bibitem[{\citenamefont{Ellis and Lewicki}(2021)}]{Ellis:2020ena}
\bibinfo{author}{\bibfnamefont{J.}~\bibnamefont{Ellis}} \bibnamefont{and}
  \bibinfo{author}{\bibfnamefont{M.}~\bibnamefont{Lewicki}},
  \bibinfo{journal}{Phys. Rev. Lett.} \textbf{\bibinfo{volume}{126}},
  \bibinfo{pages}{041304} (\bibinfo{year}{2021}), \eprint{2009.06555}.

\bibitem[{\citenamefont{Bian et~al.}(2021)\citenamefont{Bian, Cai, Liu, Yang,
  and Zhou}}]{Bian:2021lmz}
\bibinfo{author}{\bibfnamefont{L.}~\bibnamefont{Bian}},
  \bibinfo{author}{\bibfnamefont{R.-G.} \bibnamefont{Cai}},
  \bibinfo{author}{\bibfnamefont{J.}~\bibnamefont{Liu}},
  \bibinfo{author}{\bibfnamefont{X.-Y.} \bibnamefont{Yang}}, \bibnamefont{and}
  \bibinfo{author}{\bibfnamefont{R.}~\bibnamefont{Zhou}},
  \bibinfo{journal}{Phys. Rev. D} \textbf{\bibinfo{volume}{103}},
  \bibinfo{pages}{L081301} (\bibinfo{year}{2021}), \eprint{2009.13893}.

\bibitem[{\citenamefont{Ratzinger and Schwaller}(2021)}]{Ratzinger:2020koh}
\bibinfo{author}{\bibfnamefont{W.}~\bibnamefont{Ratzinger}} \bibnamefont{and}
  \bibinfo{author}{\bibfnamefont{P.}~\bibnamefont{Schwaller}},
  \bibinfo{journal}{SciPost Phys.} \textbf{\bibinfo{volume}{10}},
  \bibinfo{pages}{047} (\bibinfo{year}{2021}), \eprint{2009.11875}.

\bibitem[{\citenamefont{Addazi et~al.}(2020)\citenamefont{Addazi, Cai, Gan,
  Marciano, and Zeng}}]{Addazi:2020zcj}
\bibinfo{author}{\bibfnamefont{A.}~\bibnamefont{Addazi}},
  \bibinfo{author}{\bibfnamefont{Y.-F.} \bibnamefont{Cai}},
  \bibinfo{author}{\bibfnamefont{Q.}~\bibnamefont{Gan}},
  \bibinfo{author}{\bibfnamefont{A.}~\bibnamefont{Marciano}}, \bibnamefont{and}
  \bibinfo{author}{\bibfnamefont{K.}~\bibnamefont{Zeng}}
  (\bibinfo{year}{2020}), \eprint{2009.10327}.

\bibitem[{\citenamefont{Nakai et~al.}(2021)\citenamefont{Nakai, Suzuki,
  Takahashi, and Yamada}}]{Nakai:2020oit}
\bibinfo{author}{\bibfnamefont{Y.}~\bibnamefont{Nakai}},
  \bibinfo{author}{\bibfnamefont{M.}~\bibnamefont{Suzuki}},
  \bibinfo{author}{\bibfnamefont{F.}~\bibnamefont{Takahashi}},
  \bibnamefont{and} \bibinfo{author}{\bibfnamefont{M.}~\bibnamefont{Yamada}},
  \bibinfo{journal}{Phys. Lett. B} \textbf{\bibinfo{volume}{816}},
  \bibinfo{pages}{136238} (\bibinfo{year}{2021}), \eprint{2009.09754}.

\bibitem[{\citenamefont{Zhou et~al.}(2021)\citenamefont{Zhou, Bian, and
  Shu}}]{Zhou:2021cfu}
\bibinfo{author}{\bibfnamefont{R.}~\bibnamefont{Zhou}},
  \bibinfo{author}{\bibfnamefont{L.}~\bibnamefont{Bian}}, \bibnamefont{and}
  \bibinfo{author}{\bibfnamefont{J.}~\bibnamefont{Shu}} (\bibinfo{year}{2021}),
  \eprint{2104.03519}.

\bibitem[{\citenamefont{Borah et~al.}(2021)\citenamefont{Borah, Dasgupta, and
  Kang}}]{Borah:2021ocu}
\bibinfo{author}{\bibfnamefont{D.}~\bibnamefont{Borah}},
  \bibinfo{author}{\bibfnamefont{A.}~\bibnamefont{Dasgupta}}, \bibnamefont{and}
  \bibinfo{author}{\bibfnamefont{S.~K.} \bibnamefont{Kang}}
  (\bibinfo{year}{2021}), \eprint{2105.01007}.

\bibitem[{\citenamefont{Vagnozzi}(2021)}]{Vagnozzi:2020gtf}
\bibinfo{author}{\bibfnamefont{S.}~\bibnamefont{Vagnozzi}},
  \bibinfo{journal}{Mon. Not. Roy. Astron. Soc.}
  \textbf{\bibinfo{volume}{502}}, \bibinfo{pages}{L11} (\bibinfo{year}{2021}),
  \eprint{2009.13432}.

\bibitem[{\citenamefont{Caprini et~al.}(2016)}]{Caprini:2015zlo}
\bibinfo{author}{\bibfnamefont{C.}~\bibnamefont{Caprini}} \bibnamefont{et~al.},
  \bibinfo{journal}{JCAP} \textbf{\bibinfo{volume}{1604}}, \bibinfo{pages}{001}
  (\bibinfo{year}{2016}), \eprint{1512.06239}.

\bibitem[{\citenamefont{Caprini et~al.}(2019)}]{Caprini:2019egz}
\bibinfo{author}{\bibfnamefont{C.}~\bibnamefont{Caprini}} \bibnamefont{et~al.}
  (\bibinfo{year}{2019}), \eprint{1910.13125}.

\bibitem[{\citenamefont{Garcia-Bellido
  et~al.}(2021)\citenamefont{Garcia-Bellido, Murayama, and
  White}}]{Garcia-Bellido:2021zgu}
\bibinfo{author}{\bibfnamefont{J.}~\bibnamefont{Garcia-Bellido}},
  \bibinfo{author}{\bibfnamefont{H.}~\bibnamefont{Murayama}}, \bibnamefont{and}
  \bibinfo{author}{\bibfnamefont{G.}~\bibnamefont{White}}
  (\bibinfo{year}{2021}), \eprint{2104.04778}.

\bibitem[{\citenamefont{Goncharov et~al.}(2021)}]{Goncharov:2021oub}
\bibinfo{author}{\bibfnamefont{B.}~\bibnamefont{Goncharov}}
  \bibnamefont{et~al.} (\bibinfo{year}{2021}), \eprint{2107.12112}.

\bibitem[{\citenamefont{Jinno and
  Takimoto}(2017{\natexlab{a}})}]{Jinno:2016knw}
\bibinfo{author}{\bibfnamefont{R.}~\bibnamefont{Jinno}} \bibnamefont{and}
  \bibinfo{author}{\bibfnamefont{M.}~\bibnamefont{Takimoto}},
  \bibinfo{journal}{Phys. Rev. D} \textbf{\bibinfo{volume}{95}},
  \bibinfo{pages}{015020} (\bibinfo{year}{2017}{\natexlab{a}}),
  \eprint{1604.05035}.

\bibitem[{\citenamefont{Mohamadnejad}(2020)}]{Mohamadnejad:2019vzg}
\bibinfo{author}{\bibfnamefont{A.}~\bibnamefont{Mohamadnejad}},
  \bibinfo{journal}{Eur. Phys. J. C} \textbf{\bibinfo{volume}{80}},
  \bibinfo{pages}{197} (\bibinfo{year}{2020}), \eprint{1907.08899}.

\bibitem[{\citenamefont{Kim et~al.}(2019)\citenamefont{Kim, Lee, and
  Nam}}]{Kim:2019ogz}
\bibinfo{author}{\bibfnamefont{Y.~G.} \bibnamefont{Kim}},
  \bibinfo{author}{\bibfnamefont{K.~Y.} \bibnamefont{Lee}}, \bibnamefont{and}
  \bibinfo{author}{\bibfnamefont{S.-H.} \bibnamefont{Nam}},
  \bibinfo{journal}{Phys. Rev. D} \textbf{\bibinfo{volume}{100}},
  \bibinfo{pages}{075038} (\bibinfo{year}{2019}), \eprint{1906.03390}.

\bibitem[{\citenamefont{Hasegawa et~al.}(2019)\citenamefont{Hasegawa, Okada,
  and Seto}}]{Hasegawa:2019amx}
\bibinfo{author}{\bibfnamefont{T.}~\bibnamefont{Hasegawa}},
  \bibinfo{author}{\bibfnamefont{N.}~\bibnamefont{Okada}}, \bibnamefont{and}
  \bibinfo{author}{\bibfnamefont{O.}~\bibnamefont{Seto}},
  \bibinfo{journal}{Phys. Rev. D} \textbf{\bibinfo{volume}{99}},
  \bibinfo{pages}{095039} (\bibinfo{year}{2019}), \eprint{1904.03020}.

\bibitem[{\citenamefont{Marzo et~al.}(2019)\citenamefont{Marzo, Marzola, and
  Vaskonen}}]{Marzo:2018nov}
\bibinfo{author}{\bibfnamefont{C.}~\bibnamefont{Marzo}},
  \bibinfo{author}{\bibfnamefont{L.}~\bibnamefont{Marzola}}, \bibnamefont{and}
  \bibinfo{author}{\bibfnamefont{V.}~\bibnamefont{Vaskonen}},
  \bibinfo{journal}{Eur. Phys. J. C} \textbf{\bibinfo{volume}{79}},
  \bibinfo{pages}{601} (\bibinfo{year}{2019}), \eprint{1811.11169}.

\bibitem[{\citenamefont{Hashino et~al.}(2018)\citenamefont{Hashino, Kakizaki,
  Kanemura, Ko, and Matsui}}]{Hashino:2018zsi}
\bibinfo{author}{\bibfnamefont{K.}~\bibnamefont{Hashino}},
  \bibinfo{author}{\bibfnamefont{M.}~\bibnamefont{Kakizaki}},
  \bibinfo{author}{\bibfnamefont{S.}~\bibnamefont{Kanemura}},
  \bibinfo{author}{\bibfnamefont{P.}~\bibnamefont{Ko}}, \bibnamefont{and}
  \bibinfo{author}{\bibfnamefont{T.}~\bibnamefont{Matsui}},
  \bibinfo{journal}{JHEP} \textbf{\bibinfo{volume}{06}}, \bibinfo{pages}{088}
  (\bibinfo{year}{2018}), \eprint{1802.02947}.

\bibitem[{\citenamefont{Chiang and Senaha}(2017)}]{Chiang:2017zbz}
\bibinfo{author}{\bibfnamefont{C.-W.} \bibnamefont{Chiang}} \bibnamefont{and}
  \bibinfo{author}{\bibfnamefont{E.}~\bibnamefont{Senaha}},
  \bibinfo{journal}{Phys. Lett. B} \textbf{\bibinfo{volume}{774}},
  \bibinfo{pages}{489} (\bibinfo{year}{2017}), \eprint{1707.06765}.

\bibitem[{\citenamefont{Schwaller}(2015)}]{Schwaller:2015tja}
\bibinfo{author}{\bibfnamefont{P.}~\bibnamefont{Schwaller}},
  \bibinfo{journal}{Phys. Rev. Lett.} \textbf{\bibinfo{volume}{115}},
  \bibinfo{pages}{181101} (\bibinfo{year}{2015}), \eprint{1504.07263}.

\bibitem[{\citenamefont{Baldes and Garcia-Cely}(2019)}]{Baldes:2018emh}
\bibinfo{author}{\bibfnamefont{I.}~\bibnamefont{Baldes}} \bibnamefont{and}
  \bibinfo{author}{\bibfnamefont{C.}~\bibnamefont{Garcia-Cely}},
  \bibinfo{journal}{JHEP} \textbf{\bibinfo{volume}{05}}, \bibinfo{pages}{190}
  (\bibinfo{year}{2019}), \eprint{1809.01198}.

\bibitem[{\citenamefont{Prokopec et~al.}(2019)\citenamefont{Prokopec, Rezacek,
  and \'Swie\.zewska}}]{Prokopec:2018tnq}
\bibinfo{author}{\bibfnamefont{T.}~\bibnamefont{Prokopec}},
  \bibinfo{author}{\bibfnamefont{J.}~\bibnamefont{Rezacek}}, \bibnamefont{and}
  \bibinfo{author}{\bibfnamefont{B.}~\bibnamefont{\'Swie\.zewska}},
  \bibinfo{journal}{JCAP} \textbf{\bibinfo{volume}{02}}, \bibinfo{pages}{009}
  (\bibinfo{year}{2019}), \eprint{1809.11129}.

\bibitem[{\citenamefont{Aghanim et~al.}(2018)}]{Aghanim:2018eyx}
\bibinfo{author}{\bibfnamefont{N.}~\bibnamefont{Aghanim}} \bibnamefont{et~al.}
  (\bibinfo{collaboration}{Planck}) (\bibinfo{year}{2018}),
  \eprint{1807.06209}.

\bibitem[{\citenamefont{Schmitz}(2020)}]{Schmitz:2020syl}
\bibinfo{author}{\bibfnamefont{K.}~\bibnamefont{Schmitz}}
  (\bibinfo{year}{2020}), \eprint{2002.04615}.

\bibitem[{\citenamefont{Weltman et~al.}(2020)}]{Weltman:2018zrl}
\bibinfo{author}{\bibfnamefont{A.}~\bibnamefont{Weltman}} \bibnamefont{et~al.},
  \bibinfo{journal}{Publ. Astron. Soc. Austral.} \textbf{\bibinfo{volume}{37}},
  \bibinfo{pages}{e002} (\bibinfo{year}{2020}), \eprint{1810.02680}.

\bibitem[{\citenamefont{Hobbs et~al.}(2010)\citenamefont{Hobbs, Archibald,
  Arzoumanian, Backer, Bailes, Bhat, Burgay, Burke-Spolaor, Champion, Cognard
  et~al.}}]{Hobbs_2010}
\bibinfo{author}{\bibfnamefont{G.}~\bibnamefont{Hobbs}},
  \bibinfo{author}{\bibfnamefont{A.}~\bibnamefont{Archibald}},
  \bibinfo{author}{\bibfnamefont{Z.}~\bibnamefont{Arzoumanian}},
  \bibinfo{author}{\bibfnamefont{D.}~\bibnamefont{Backer}},
  \bibinfo{author}{\bibfnamefont{M.}~\bibnamefont{Bailes}},
  \bibinfo{author}{\bibfnamefont{N.~D.~R.} \bibnamefont{Bhat}},
  \bibinfo{author}{\bibfnamefont{M.}~\bibnamefont{Burgay}},
  \bibinfo{author}{\bibfnamefont{S.}~\bibnamefont{Burke-Spolaor}},
  \bibinfo{author}{\bibfnamefont{D.}~\bibnamefont{Champion}},
  \bibinfo{author}{\bibfnamefont{I.}~\bibnamefont{Cognard}},
  \bibnamefont{et~al.}, \bibinfo{journal}{Classical and Quantum Gravity}
  \textbf{\bibinfo{volume}{27}}, \bibinfo{pages}{084013}
  (\bibinfo{year}{2010}), ISSN \bibinfo{issn}{1361-6382},
  \urlprefix\url{http://dx.doi.org/10.1088/0264-9381/27/8/084013}.

\bibitem[{\citenamefont{Dolan and Jackiw}(1974)}]{Dolan:1973qd}
\bibinfo{author}{\bibfnamefont{L.}~\bibnamefont{Dolan}} \bibnamefont{and}
  \bibinfo{author}{\bibfnamefont{R.}~\bibnamefont{Jackiw}},
  \bibinfo{journal}{Phys. Rev.} \textbf{\bibinfo{volume}{D9}},
  \bibinfo{pages}{3320} (\bibinfo{year}{1974}).

\bibitem[{\citenamefont{Quiros}(1999)}]{Quiros:1999jp}
\bibinfo{author}{\bibfnamefont{M.}~\bibnamefont{Quiros}}, in
  \emph{\bibinfo{booktitle}{{Proceedings, Summer School in High-energy physics
  and cosmology: Trieste, Italy, June 29-July 17, 1998}}}
  (\bibinfo{year}{1999}), pp. \bibinfo{pages}{187--259},
  \eprint{hep-ph/9901312}.

\bibitem[{\citenamefont{Fendley}(1987)}]{Fendley:1987ef}
\bibinfo{author}{\bibfnamefont{P.}~\bibnamefont{Fendley}},
  \bibinfo{journal}{Phys. Lett.} \textbf{\bibinfo{volume}{B196}},
  \bibinfo{pages}{175} (\bibinfo{year}{1987}).

\bibitem[{\citenamefont{Parwani}(1992)}]{Parwani:1991gq}
\bibinfo{author}{\bibfnamefont{R.~R.} \bibnamefont{Parwani}},
  \bibinfo{journal}{Phys. Rev.} \textbf{\bibinfo{volume}{D45}},
  \bibinfo{pages}{4695} (\bibinfo{year}{1992}), \bibinfo{note}{[Erratum: Phys.
  Rev.D48,5965(1993)]}, \eprint{hep-ph/9204216}.

\bibitem[{\citenamefont{Arnold and Espinosa}(1993)}]{Arnold:1992rz}
\bibinfo{author}{\bibfnamefont{P.~B.} \bibnamefont{Arnold}} \bibnamefont{and}
  \bibinfo{author}{\bibfnamefont{O.}~\bibnamefont{Espinosa}},
  \bibinfo{journal}{Phys. Rev.} \textbf{\bibinfo{volume}{D47}},
  \bibinfo{pages}{3546} (\bibinfo{year}{1993}), \bibinfo{note}{[Erratum: Phys.
  Rev.D50,6662(1994)]}, \eprint{hep-ph/9212235}.

\bibitem[{\citenamefont{Cline et~al.}(2008)\citenamefont{Cline, Jarvinen, and
  Sannino}}]{Cline:2008hr}
\bibinfo{author}{\bibfnamefont{J.~M.} \bibnamefont{Cline}},
  \bibinfo{author}{\bibfnamefont{M.}~\bibnamefont{Jarvinen}}, \bibnamefont{and}
  \bibinfo{author}{\bibfnamefont{F.}~\bibnamefont{Sannino}},
  \bibinfo{journal}{Phys. Rev. D} \textbf{\bibinfo{volume}{78}},
  \bibinfo{pages}{075027} (\bibinfo{year}{2008}), \eprint{0808.1512}.

\bibitem[{\citenamefont{Mazumdar and White}(2019)}]{Mazumdar:2018dfl}
\bibinfo{author}{\bibfnamefont{A.}~\bibnamefont{Mazumdar}} \bibnamefont{and}
  \bibinfo{author}{\bibfnamefont{G.}~\bibnamefont{White}},
  \bibinfo{journal}{Rept. Prog. Phys.} \textbf{\bibinfo{volume}{82}},
  \bibinfo{pages}{076901} (\bibinfo{year}{2019}), \eprint{1811.01948}.

\bibitem[{\citenamefont{Hindmarsh et~al.}(2021)\citenamefont{Hindmarsh,
  L\"uben, Lumma, and Pauly}}]{Hindmarsh:2020hop}
\bibinfo{author}{\bibfnamefont{M.~B.} \bibnamefont{Hindmarsh}},
  \bibinfo{author}{\bibfnamefont{M.}~\bibnamefont{L\"uben}},
  \bibinfo{author}{\bibfnamefont{J.}~\bibnamefont{Lumma}}, \bibnamefont{and}
  \bibinfo{author}{\bibfnamefont{M.}~\bibnamefont{Pauly}},
  \bibinfo{journal}{SciPost Phys. Lect. Notes} \textbf{\bibinfo{volume}{24}},
  \bibinfo{pages}{1} (\bibinfo{year}{2021}), \eprint{2008.09136}.

\bibitem[{\citenamefont{Linde}(1983)}]{Linde:1981zj}
\bibinfo{author}{\bibfnamefont{A.~D.} \bibnamefont{Linde}},
  \bibinfo{journal}{Nucl. Phys. B} \textbf{\bibinfo{volume}{216}},
  \bibinfo{pages}{421} (\bibinfo{year}{1983}), \bibinfo{note}{[Erratum:
  Nucl.Phys.B 223, 544 (1983)]}.

\bibitem[{\citenamefont{Linde}(1981)}]{Linde:1980tt}
\bibinfo{author}{\bibfnamefont{A.~D.} \bibnamefont{Linde}},
  \bibinfo{journal}{Phys. Lett.} \textbf{\bibinfo{volume}{100B}},
  \bibinfo{pages}{37} (\bibinfo{year}{1981}).

\bibitem[{\citenamefont{Borah et~al.}(2020)\citenamefont{Borah, Dasgupta,
  Fujikura, Kang, and Mahanta}}]{Borah:2020wut}
\bibinfo{author}{\bibfnamefont{D.}~\bibnamefont{Borah}},
  \bibinfo{author}{\bibfnamefont{A.}~\bibnamefont{Dasgupta}},
  \bibinfo{author}{\bibfnamefont{K.}~\bibnamefont{Fujikura}},
  \bibinfo{author}{\bibfnamefont{S.~K.} \bibnamefont{Kang}}, \bibnamefont{and}
  \bibinfo{author}{\bibfnamefont{D.}~\bibnamefont{Mahanta}},
  \bibinfo{journal}{JCAP} \textbf{\bibinfo{volume}{08}}, \bibinfo{pages}{046}
  (\bibinfo{year}{2020}), \eprint{2003.02276}.

\bibitem[{\citenamefont{Ellis et~al.}(2018)\citenamefont{Ellis, Lewicki, and
  No}}]{Ellis:2018mja}
\bibinfo{author}{\bibfnamefont{J.}~\bibnamefont{Ellis}},
  \bibinfo{author}{\bibfnamefont{M.}~\bibnamefont{Lewicki}}, \bibnamefont{and}
  \bibinfo{author}{\bibfnamefont{J.~M.} \bibnamefont{No}}
  (\bibinfo{year}{2018}), \bibinfo{note}{[JCAP1904,003(2019)]},
  \eprint{1809.08242}.

\bibitem[{\citenamefont{Ellis et~al.}(2020)\citenamefont{Ellis, Lewicki, and
  Vaskonen}}]{Ellis:2020nnr}
\bibinfo{author}{\bibfnamefont{J.}~\bibnamefont{Ellis}},
  \bibinfo{author}{\bibfnamefont{M.}~\bibnamefont{Lewicki}}, \bibnamefont{and}
  \bibinfo{author}{\bibfnamefont{V.}~\bibnamefont{Vaskonen}},
  \bibinfo{journal}{JCAP} \textbf{\bibinfo{volume}{11}}, \bibinfo{pages}{020}
  (\bibinfo{year}{2020}), \eprint{2007.15586}.

\bibitem[{\citenamefont{Turner and Wilczek}(1990)}]{Turner:1990rc}
\bibinfo{author}{\bibfnamefont{M.~S.} \bibnamefont{Turner}} \bibnamefont{and}
  \bibinfo{author}{\bibfnamefont{F.}~\bibnamefont{Wilczek}},
  \bibinfo{journal}{Phys. Rev. Lett.} \textbf{\bibinfo{volume}{65}},
  \bibinfo{pages}{3080} (\bibinfo{year}{1990}).

\bibitem[{\citenamefont{Kosowsky
  et~al.}(1992{\natexlab{a}})\citenamefont{Kosowsky, Turner, and
  Watkins}}]{Kosowsky:1991ua}
\bibinfo{author}{\bibfnamefont{A.}~\bibnamefont{Kosowsky}},
  \bibinfo{author}{\bibfnamefont{M.~S.} \bibnamefont{Turner}},
  \bibnamefont{and} \bibinfo{author}{\bibfnamefont{R.}~\bibnamefont{Watkins}},
  \bibinfo{journal}{Phys. Rev.} \textbf{\bibinfo{volume}{D45}},
  \bibinfo{pages}{4514} (\bibinfo{year}{1992}{\natexlab{a}}).

\bibitem[{\citenamefont{Kosowsky
  et~al.}(1992{\natexlab{b}})\citenamefont{Kosowsky, Turner, and
  Watkins}}]{Kosowsky:1992rz}
\bibinfo{author}{\bibfnamefont{A.}~\bibnamefont{Kosowsky}},
  \bibinfo{author}{\bibfnamefont{M.~S.} \bibnamefont{Turner}},
  \bibnamefont{and} \bibinfo{author}{\bibfnamefont{R.}~\bibnamefont{Watkins}},
  \bibinfo{journal}{Phys. Rev. Lett.} \textbf{\bibinfo{volume}{69}},
  \bibinfo{pages}{2026} (\bibinfo{year}{1992}{\natexlab{b}}).

\bibitem[{\citenamefont{Kosowsky and Turner}(1993)}]{Kosowsky:1992vn}
\bibinfo{author}{\bibfnamefont{A.}~\bibnamefont{Kosowsky}} \bibnamefont{and}
  \bibinfo{author}{\bibfnamefont{M.~S.} \bibnamefont{Turner}},
  \bibinfo{journal}{Phys. Rev.} \textbf{\bibinfo{volume}{D47}},
  \bibinfo{pages}{4372} (\bibinfo{year}{1993}), \eprint{astro-ph/9211004}.

\bibitem[{\citenamefont{Turner et~al.}(1992)\citenamefont{Turner, Weinberg, and
  Widrow}}]{Turner:1992tz}
\bibinfo{author}{\bibfnamefont{M.~S.} \bibnamefont{Turner}},
  \bibinfo{author}{\bibfnamefont{E.~J.} \bibnamefont{Weinberg}},
  \bibnamefont{and} \bibinfo{author}{\bibfnamefont{L.~M.}
  \bibnamefont{Widrow}}, \bibinfo{journal}{Phys. Rev.}
  \textbf{\bibinfo{volume}{D46}}, \bibinfo{pages}{2384} (\bibinfo{year}{1992}).

\bibitem[{\citenamefont{Hindmarsh et~al.}(2014)\citenamefont{Hindmarsh, Huber,
  Rummukainen, and Weir}}]{Hindmarsh:2013xza}
\bibinfo{author}{\bibfnamefont{M.}~\bibnamefont{Hindmarsh}},
  \bibinfo{author}{\bibfnamefont{S.~J.} \bibnamefont{Huber}},
  \bibinfo{author}{\bibfnamefont{K.}~\bibnamefont{Rummukainen}},
  \bibnamefont{and} \bibinfo{author}{\bibfnamefont{D.~J.} \bibnamefont{Weir}},
  \bibinfo{journal}{Phys. Rev. Lett.} \textbf{\bibinfo{volume}{112}},
  \bibinfo{pages}{041301} (\bibinfo{year}{2014}), \eprint{1304.2433}.

\bibitem[{\citenamefont{Giblin and Mertens}(2014)}]{Giblin:2014qia}
\bibinfo{author}{\bibfnamefont{J.~T.} \bibnamefont{Giblin}} \bibnamefont{and}
  \bibinfo{author}{\bibfnamefont{J.~B.} \bibnamefont{Mertens}},
  \bibinfo{journal}{Phys. Rev.} \textbf{\bibinfo{volume}{D90}},
  \bibinfo{pages}{023532} (\bibinfo{year}{2014}), \eprint{1405.4005}.

\bibitem[{\citenamefont{Hindmarsh et~al.}(2015)\citenamefont{Hindmarsh, Huber,
  Rummukainen, and Weir}}]{Hindmarsh:2015qta}
\bibinfo{author}{\bibfnamefont{M.}~\bibnamefont{Hindmarsh}},
  \bibinfo{author}{\bibfnamefont{S.~J.} \bibnamefont{Huber}},
  \bibinfo{author}{\bibfnamefont{K.}~\bibnamefont{Rummukainen}},
  \bibnamefont{and} \bibinfo{author}{\bibfnamefont{D.~J.} \bibnamefont{Weir}},
  \bibinfo{journal}{Phys. Rev.} \textbf{\bibinfo{volume}{D92}},
  \bibinfo{pages}{123009} (\bibinfo{year}{2015}), \eprint{1504.03291}.

\bibitem[{\citenamefont{Hindmarsh et~al.}(2017)\citenamefont{Hindmarsh, Huber,
  Rummukainen, and Weir}}]{Hindmarsh:2017gnf}
\bibinfo{author}{\bibfnamefont{M.}~\bibnamefont{Hindmarsh}},
  \bibinfo{author}{\bibfnamefont{S.~J.} \bibnamefont{Huber}},
  \bibinfo{author}{\bibfnamefont{K.}~\bibnamefont{Rummukainen}},
  \bibnamefont{and} \bibinfo{author}{\bibfnamefont{D.~J.} \bibnamefont{Weir}},
  \bibinfo{journal}{Phys. Rev.} \textbf{\bibinfo{volume}{D96}},
  \bibinfo{pages}{103520} (\bibinfo{year}{2017}), \eprint{1704.05871}.

\bibitem[{\citenamefont{Kamionkowski et~al.}(1994)\citenamefont{Kamionkowski,
  Kosowsky, and Turner}}]{Kamionkowski:1993fg}
\bibinfo{author}{\bibfnamefont{M.}~\bibnamefont{Kamionkowski}},
  \bibinfo{author}{\bibfnamefont{A.}~\bibnamefont{Kosowsky}}, \bibnamefont{and}
  \bibinfo{author}{\bibfnamefont{M.~S.} \bibnamefont{Turner}},
  \bibinfo{journal}{Phys. Rev.} \textbf{\bibinfo{volume}{D49}},
  \bibinfo{pages}{2837} (\bibinfo{year}{1994}), \eprint{astro-ph/9310044}.

\bibitem[{\citenamefont{Kosowsky et~al.}(2002)\citenamefont{Kosowsky, Mack, and
  Kahniashvili}}]{Kosowsky:2001xp}
\bibinfo{author}{\bibfnamefont{A.}~\bibnamefont{Kosowsky}},
  \bibinfo{author}{\bibfnamefont{A.}~\bibnamefont{Mack}}, \bibnamefont{and}
  \bibinfo{author}{\bibfnamefont{T.}~\bibnamefont{Kahniashvili}},
  \bibinfo{journal}{Phys. Rev.} \textbf{\bibinfo{volume}{D66}},
  \bibinfo{pages}{024030} (\bibinfo{year}{2002}), \eprint{astro-ph/0111483}.

\bibitem[{\citenamefont{Caprini and Durrer}(2006)}]{Caprini:2006jb}
\bibinfo{author}{\bibfnamefont{C.}~\bibnamefont{Caprini}} \bibnamefont{and}
  \bibinfo{author}{\bibfnamefont{R.}~\bibnamefont{Durrer}},
  \bibinfo{journal}{Phys. Rev.} \textbf{\bibinfo{volume}{D74}},
  \bibinfo{pages}{063521} (\bibinfo{year}{2006}), \eprint{astro-ph/0603476}.

\bibitem[{\citenamefont{Gogoberidze et~al.}(2007)\citenamefont{Gogoberidze,
  Kahniashvili, and Kosowsky}}]{Gogoberidze:2007an}
\bibinfo{author}{\bibfnamefont{G.}~\bibnamefont{Gogoberidze}},
  \bibinfo{author}{\bibfnamefont{T.}~\bibnamefont{Kahniashvili}},
  \bibnamefont{and} \bibinfo{author}{\bibfnamefont{A.}~\bibnamefont{Kosowsky}},
  \bibinfo{journal}{Phys. Rev.} \textbf{\bibinfo{volume}{D76}},
  \bibinfo{pages}{083002} (\bibinfo{year}{2007}), \eprint{0705.1733}.

\bibitem[{\citenamefont{Caprini et~al.}(2009)\citenamefont{Caprini, Durrer, and
  Servant}}]{Caprini:2009yp}
\bibinfo{author}{\bibfnamefont{C.}~\bibnamefont{Caprini}},
  \bibinfo{author}{\bibfnamefont{R.}~\bibnamefont{Durrer}}, \bibnamefont{and}
  \bibinfo{author}{\bibfnamefont{G.}~\bibnamefont{Servant}},
  \bibinfo{journal}{JCAP} \textbf{\bibinfo{volume}{0912}}, \bibinfo{pages}{024}
  (\bibinfo{year}{2009}), \eprint{0909.0622}.

\bibitem[{\citenamefont{Niksa et~al.}(2018)\citenamefont{Niksa, Schlederer, and
  Sigl}}]{Niksa:2018ofa}
\bibinfo{author}{\bibfnamefont{P.}~\bibnamefont{Niksa}},
  \bibinfo{author}{\bibfnamefont{M.}~\bibnamefont{Schlederer}},
  \bibnamefont{and} \bibinfo{author}{\bibfnamefont{G.}~\bibnamefont{Sigl}},
  \bibinfo{journal}{Class. Quant. Grav.} \textbf{\bibinfo{volume}{35}},
  \bibinfo{pages}{144001} (\bibinfo{year}{2018}), \eprint{1803.02271}.

\bibitem[{\citenamefont{Sato}(2021)}]{Sato:2019wpo}
\bibinfo{author}{\bibfnamefont{R.}~\bibnamefont{Sato}},
  \bibinfo{journal}{Comput. Phys. Commun.} \textbf{\bibinfo{volume}{258}},
  \bibinfo{pages}{107566} (\bibinfo{year}{2021}), \eprint{1908.10868}.

\bibitem[{\citenamefont{Athron et~al.}(2019)\citenamefont{Athron, Bal\'azs,
  Bardsley, Fowlie, Harries, and White}}]{Athron:2019nbd}
\bibinfo{author}{\bibfnamefont{P.}~\bibnamefont{Athron}},
  \bibinfo{author}{\bibfnamefont{C.}~\bibnamefont{Bal\'azs}},
  \bibinfo{author}{\bibfnamefont{M.}~\bibnamefont{Bardsley}},
  \bibinfo{author}{\bibfnamefont{A.}~\bibnamefont{Fowlie}},
  \bibinfo{author}{\bibfnamefont{D.}~\bibnamefont{Harries}}, \bibnamefont{and}
  \bibinfo{author}{\bibfnamefont{G.}~\bibnamefont{White}},
  \bibinfo{journal}{Comput. Phys. Commun.} \textbf{\bibinfo{volume}{244}},
  \bibinfo{pages}{448} (\bibinfo{year}{2019}), \eprint{1901.03714}.

\bibitem[{\citenamefont{Jinno and
  Takimoto}(2017{\natexlab{b}})}]{Jinno:2016vai}
\bibinfo{author}{\bibfnamefont{R.}~\bibnamefont{Jinno}} \bibnamefont{and}
  \bibinfo{author}{\bibfnamefont{M.}~\bibnamefont{Takimoto}},
  \bibinfo{journal}{Phys. Rev. D} \textbf{\bibinfo{volume}{95}},
  \bibinfo{pages}{024009} (\bibinfo{year}{2017}{\natexlab{b}}),
  \eprint{1605.01403}.

\bibitem[{\citenamefont{Lewicki and Vaskonen}(2020)}]{Lewicki:2020azd}
\bibinfo{author}{\bibfnamefont{M.}~\bibnamefont{Lewicki}} \bibnamefont{and}
  \bibinfo{author}{\bibfnamefont{V.}~\bibnamefont{Vaskonen}}
  (\bibinfo{year}{2020}), \eprint{2012.07826}.

\bibitem[{\citenamefont{Binetruy et~al.}(2012)\citenamefont{Binetruy, Bohe,
  Caprini, and Dufaux}}]{Binetruy:2012ze}
\bibinfo{author}{\bibfnamefont{P.}~\bibnamefont{Binetruy}},
  \bibinfo{author}{\bibfnamefont{A.}~\bibnamefont{Bohe}},
  \bibinfo{author}{\bibfnamefont{C.}~\bibnamefont{Caprini}}, \bibnamefont{and}
  \bibinfo{author}{\bibfnamefont{J.-F.} \bibnamefont{Dufaux}},
  \bibinfo{journal}{JCAP} \textbf{\bibinfo{volume}{1206}}, \bibinfo{pages}{027}
  (\bibinfo{year}{2012}), \eprint{1201.0983}.

\bibitem[{\citenamefont{Espinosa et~al.}(2010)\citenamefont{Espinosa,
  Konstandin, No, and Servant}}]{Espinosa:2010hh}
\bibinfo{author}{\bibfnamefont{J.~R.} \bibnamefont{Espinosa}},
  \bibinfo{author}{\bibfnamefont{T.}~\bibnamefont{Konstandin}},
  \bibinfo{author}{\bibfnamefont{J.~M.} \bibnamefont{No}}, \bibnamefont{and}
  \bibinfo{author}{\bibfnamefont{G.}~\bibnamefont{Servant}},
  \bibinfo{journal}{JCAP} \textbf{\bibinfo{volume}{06}}, \bibinfo{pages}{028}
  (\bibinfo{year}{2010}), \eprint{1004.4187}.

\bibitem[{\citenamefont{Steinhardt}(1982)}]{Steinhardt:1981ct}
\bibinfo{author}{\bibfnamefont{P.~J.} \bibnamefont{Steinhardt}},
  \bibinfo{journal}{Phys. Rev.} \textbf{\bibinfo{volume}{D25}},
  \bibinfo{pages}{2074} (\bibinfo{year}{1982}).

\bibitem[{\citenamefont{Huber and Sopena}(2013)}]{Huber:2013kj}
\bibinfo{author}{\bibfnamefont{S.~J.} \bibnamefont{Huber}} \bibnamefont{and}
  \bibinfo{author}{\bibfnamefont{M.}~\bibnamefont{Sopena}}
  (\bibinfo{year}{2013}), \eprint{1302.1044}.

\bibitem[{\citenamefont{Leitao and Megevand}(2015)}]{Leitao:2014pda}
\bibinfo{author}{\bibfnamefont{L.}~\bibnamefont{Leitao}} \bibnamefont{and}
  \bibinfo{author}{\bibfnamefont{A.}~\bibnamefont{Megevand}},
  \bibinfo{journal}{Nucl. Phys.} \textbf{\bibinfo{volume}{B891}},
  \bibinfo{pages}{159} (\bibinfo{year}{2015}), \eprint{1410.3875}.

\bibitem[{\citenamefont{Dorsch et~al.}(2018)\citenamefont{Dorsch, Huber, and
  Konstandin}}]{Dorsch:2018pat}
\bibinfo{author}{\bibfnamefont{G.~C.} \bibnamefont{Dorsch}},
  \bibinfo{author}{\bibfnamefont{S.~J.} \bibnamefont{Huber}}, \bibnamefont{and}
  \bibinfo{author}{\bibfnamefont{T.}~\bibnamefont{Konstandin}},
  \bibinfo{journal}{JCAP} \textbf{\bibinfo{volume}{1812}}, \bibinfo{pages}{034}
  (\bibinfo{year}{2018}), \eprint{1809.04907}.

\bibitem[{\citenamefont{Cline and Kainulainen}(2020)}]{Cline:2020jre}
\bibinfo{author}{\bibfnamefont{J.~M.} \bibnamefont{Cline}} \bibnamefont{and}
  \bibinfo{author}{\bibfnamefont{K.}~\bibnamefont{Kainulainen}}
  (\bibinfo{year}{2020}), \eprint{2001.00568}.

\bibitem[{\citenamefont{Azatov and Vanvlasselaer}(2021)}]{Azatov:2020ufh}
\bibinfo{author}{\bibfnamefont{A.}~\bibnamefont{Azatov}} \bibnamefont{and}
  \bibinfo{author}{\bibfnamefont{M.}~\bibnamefont{Vanvlasselaer}},
  \bibinfo{journal}{JCAP} \textbf{\bibinfo{volume}{01}}, \bibinfo{pages}{058}
  (\bibinfo{year}{2021}), \eprint{2010.02590}.

\bibitem[{\citenamefont{Bodeker and Moore}(2017)}]{Bodeker:2017cim}
\bibinfo{author}{\bibfnamefont{D.}~\bibnamefont{Bodeker}} \bibnamefont{and}
  \bibinfo{author}{\bibfnamefont{G.~D.} \bibnamefont{Moore}},
  \bibinfo{journal}{JCAP} \textbf{\bibinfo{volume}{1705}}, \bibinfo{pages}{025}
  (\bibinfo{year}{2017}), \eprint{1703.08215}.

\bibitem[{\citenamefont{Ellis et~al.}(2019)\citenamefont{Ellis, Lewicki, No,
  and Vaskonen}}]{Ellis:2019oqb}
\bibinfo{author}{\bibfnamefont{J.}~\bibnamefont{Ellis}},
  \bibinfo{author}{\bibfnamefont{M.}~\bibnamefont{Lewicki}},
  \bibinfo{author}{\bibfnamefont{J.~M.} \bibnamefont{No}}, \bibnamefont{and}
  \bibinfo{author}{\bibfnamefont{V.}~\bibnamefont{Vaskonen}},
  \bibinfo{journal}{JCAP} \textbf{\bibinfo{volume}{1906}}, \bibinfo{pages}{024}
  (\bibinfo{year}{2019}), \eprint{1903.09642}.

\bibitem[{\citenamefont{McLaughlin}(2013)}]{McLaughlin:2013ira}
\bibinfo{author}{\bibfnamefont{M.~A.} \bibnamefont{McLaughlin}},
  \bibinfo{journal}{Class. Quant. Grav.} \textbf{\bibinfo{volume}{30}},
  \bibinfo{pages}{224008} (\bibinfo{year}{2013}), \eprint{1310.0758}.

\bibitem[{\citenamefont{Manchester et~al.}(2013)\citenamefont{Manchester,
  Hobbs, Bailes, Coles, van Straten, Keith, Shannon, Bhat, Brown, Burke-Spolaor
  et~al.}}]{Manchester_2013}
\bibinfo{author}{\bibfnamefont{R.~N.} \bibnamefont{Manchester}},
  \bibinfo{author}{\bibfnamefont{G.}~\bibnamefont{Hobbs}},
  \bibinfo{author}{\bibfnamefont{M.}~\bibnamefont{Bailes}},
  \bibinfo{author}{\bibfnamefont{W.~A.} \bibnamefont{Coles}},
  \bibinfo{author}{\bibfnamefont{W.}~\bibnamefont{van Straten}},
  \bibinfo{author}{\bibfnamefont{M.~J.} \bibnamefont{Keith}},
  \bibinfo{author}{\bibfnamefont{R.~M.} \bibnamefont{Shannon}},
  \bibinfo{author}{\bibfnamefont{N.~D.~R.} \bibnamefont{Bhat}},
  \bibinfo{author}{\bibfnamefont{A.}~\bibnamefont{Brown}},
  \bibinfo{author}{\bibfnamefont{S.~G.} \bibnamefont{Burke-Spolaor}},
  \bibnamefont{et~al.}, \bibinfo{journal}{Publications of the Astronomical
  Society of Australia} \textbf{\bibinfo{volume}{30}} (\bibinfo{year}{2013}),
  ISSN \bibinfo{issn}{1448-6083},
  \urlprefix\url{http://dx.doi.org/10.1017/pasa.2012.017}.

\bibitem[{\citenamefont{Kramer and Champion}(2013)}]{Kramer_2013}
\bibinfo{author}{\bibfnamefont{M.}~\bibnamefont{Kramer}} \bibnamefont{and}
  \bibinfo{author}{\bibfnamefont{D.~J.} \bibnamefont{Champion}},
  \bibinfo{journal}{Classical and Quantum Gravity}
  \textbf{\bibinfo{volume}{30}}, \bibinfo{pages}{224009}
  (\bibinfo{year}{2013}),
  \urlprefix\url{https://doi.org/10.1088/0264-9381/30/22/224009}.

\bibitem[{\citenamefont{Zwicky}(1933)}]{Zwicky:1933gu}
\bibinfo{author}{\bibfnamefont{F.}~\bibnamefont{Zwicky}},
  \bibinfo{journal}{Helv. Phys. Acta} \textbf{\bibinfo{volume}{6}},
  \bibinfo{pages}{110} (\bibinfo{year}{1933}), \bibinfo{note}{[Gen. Rel.
  Grav.41,207(2009)]}.

\bibitem[{\citenamefont{Rubin and Ford}(1970)}]{Rubin:1970zza}
\bibinfo{author}{\bibfnamefont{V.~C.} \bibnamefont{Rubin}} \bibnamefont{and}
  \bibinfo{author}{\bibfnamefont{W.~K.} \bibnamefont{Ford},
  \bibfnamefont{Jr.}}, \bibinfo{journal}{Astrophys. J.}
  \textbf{\bibinfo{volume}{159}}, \bibinfo{pages}{379} (\bibinfo{year}{1970}).

\bibitem[{\citenamefont{Clowe et~al.}(2006)\citenamefont{Clowe, Bradac,
  Gonzalez, Markevitch, Randall, Jones, and Zaritsky}}]{Clowe:2006eq}
\bibinfo{author}{\bibfnamefont{D.}~\bibnamefont{Clowe}},
  \bibinfo{author}{\bibfnamefont{M.}~\bibnamefont{Bradac}},
  \bibinfo{author}{\bibfnamefont{A.~H.} \bibnamefont{Gonzalez}},
  \bibinfo{author}{\bibfnamefont{M.}~\bibnamefont{Markevitch}},
  \bibinfo{author}{\bibfnamefont{S.~W.} \bibnamefont{Randall}},
  \bibinfo{author}{\bibfnamefont{C.}~\bibnamefont{Jones}}, \bibnamefont{and}
  \bibinfo{author}{\bibfnamefont{D.}~\bibnamefont{Zaritsky}},
  \bibinfo{journal}{Astrophys. J.} \textbf{\bibinfo{volume}{648}},
  \bibinfo{pages}{L109} (\bibinfo{year}{2006}), \eprint{astro-ph/0608407}.

\bibitem[{\citenamefont{Kolb and Turner}(1990)}]{Kolb:1990vq}
\bibinfo{author}{\bibfnamefont{E.~W.} \bibnamefont{Kolb}} \bibnamefont{and}
  \bibinfo{author}{\bibfnamefont{M.~S.} \bibnamefont{Turner}},
  \bibinfo{journal}{Front. Phys.} \textbf{\bibinfo{volume}{69}},
  \bibinfo{pages}{1} (\bibinfo{year}{1990}).

\bibitem[{\citenamefont{Arcadi et~al.}(2017)\citenamefont{Arcadi, Dutra, Ghosh,
  Lindner, Mambrini, Pierre, Profumo, and Queiroz}}]{Arcadi:2017kky}
\bibinfo{author}{\bibfnamefont{G.}~\bibnamefont{Arcadi}},
  \bibinfo{author}{\bibfnamefont{M.}~\bibnamefont{Dutra}},
  \bibinfo{author}{\bibfnamefont{P.}~\bibnamefont{Ghosh}},
  \bibinfo{author}{\bibfnamefont{M.}~\bibnamefont{Lindner}},
  \bibinfo{author}{\bibfnamefont{Y.}~\bibnamefont{Mambrini}},
  \bibinfo{author}{\bibfnamefont{M.}~\bibnamefont{Pierre}},
  \bibinfo{author}{\bibfnamefont{S.}~\bibnamefont{Profumo}}, \bibnamefont{and}
  \bibinfo{author}{\bibfnamefont{F.~S.} \bibnamefont{Queiroz}}
  (\bibinfo{year}{2017}), \eprint{1703.07364}.

\bibitem[{\citenamefont{Hall et~al.}(2010)\citenamefont{Hall, Jedamzik,
  March-Russell, and West}}]{Hall:2009bx}
\bibinfo{author}{\bibfnamefont{L.~J.} \bibnamefont{Hall}},
  \bibinfo{author}{\bibfnamefont{K.}~\bibnamefont{Jedamzik}},
  \bibinfo{author}{\bibfnamefont{J.}~\bibnamefont{March-Russell}},
  \bibnamefont{and} \bibinfo{author}{\bibfnamefont{S.~M.} \bibnamefont{West}},
  \bibinfo{journal}{JHEP} \textbf{\bibinfo{volume}{03}}, \bibinfo{pages}{080}
  (\bibinfo{year}{2010}), \eprint{0911.1120}.

\bibitem[{\citenamefont{Bernal et~al.}(2017)\citenamefont{Bernal, Heikinheimo,
  Tenkanen, Tuominen, and Vaskonen}}]{Bernal:2017kxu}
\bibinfo{author}{\bibfnamefont{N.}~\bibnamefont{Bernal}},
  \bibinfo{author}{\bibfnamefont{M.}~\bibnamefont{Heikinheimo}},
  \bibinfo{author}{\bibfnamefont{T.}~\bibnamefont{Tenkanen}},
  \bibinfo{author}{\bibfnamefont{K.}~\bibnamefont{Tuominen}}, \bibnamefont{and}
  \bibinfo{author}{\bibfnamefont{V.}~\bibnamefont{Vaskonen}},
  \bibinfo{journal}{Int. J. Mod. Phys.} \textbf{\bibinfo{volume}{A32}},
  \bibinfo{pages}{1730023} (\bibinfo{year}{2017}), \eprint{1706.07442}.

\bibitem[{\citenamefont{Diaz-Cruz and Ma}(2011)}]{Diaz-Cruz:2010czr}
\bibinfo{author}{\bibfnamefont{J.~L.} \bibnamefont{Diaz-Cruz}}
  \bibnamefont{and} \bibinfo{author}{\bibfnamefont{E.}~\bibnamefont{Ma}},
  \bibinfo{journal}{Phys. Lett. B} \textbf{\bibinfo{volume}{695}},
  \bibinfo{pages}{264} (\bibinfo{year}{2011}), \eprint{1007.2631}.

\bibitem[{\citenamefont{Fraser et~al.}(2015)\citenamefont{Fraser, Ma, and
  Zakeri}}]{Fraser:2014yga}
\bibinfo{author}{\bibfnamefont{S.}~\bibnamefont{Fraser}},
  \bibinfo{author}{\bibfnamefont{E.}~\bibnamefont{Ma}}, \bibnamefont{and}
  \bibinfo{author}{\bibfnamefont{M.}~\bibnamefont{Zakeri}},
  \bibinfo{journal}{Int. J. Mod. Phys. A} \textbf{\bibinfo{volume}{30}},
  \bibinfo{pages}{1550018} (\bibinfo{year}{2015}), \eprint{1409.1162}.

\bibitem[{\citenamefont{Bhattacharya et~al.}(2012)\citenamefont{Bhattacharya,
  Diaz-Cruz, Ma, and Wegman}}]{Bhattacharya:2011tr}
\bibinfo{author}{\bibfnamefont{S.}~\bibnamefont{Bhattacharya}},
  \bibinfo{author}{\bibfnamefont{J.~L.} \bibnamefont{Diaz-Cruz}},
  \bibinfo{author}{\bibfnamefont{E.}~\bibnamefont{Ma}}, \bibnamefont{and}
  \bibinfo{author}{\bibfnamefont{D.}~\bibnamefont{Wegman}},
  \bibinfo{journal}{Phys. Rev. D} \textbf{\bibinfo{volume}{85}},
  \bibinfo{pages}{055008} (\bibinfo{year}{2012}), \eprint{1107.2093}.

\bibitem[{\citenamefont{Barman et~al.}(2017)\citenamefont{Barman, Bhattacharya,
  Patra, and Chakrabortty}}]{Barman:2017yzr}
\bibinfo{author}{\bibfnamefont{B.}~\bibnamefont{Barman}},
  \bibinfo{author}{\bibfnamefont{S.}~\bibnamefont{Bhattacharya}},
  \bibinfo{author}{\bibfnamefont{S.~K.} \bibnamefont{Patra}}, \bibnamefont{and}
  \bibinfo{author}{\bibfnamefont{J.}~\bibnamefont{Chakrabortty}},
  \bibinfo{journal}{JCAP} \textbf{\bibinfo{volume}{12}}, \bibinfo{pages}{021}
  (\bibinfo{year}{2017}), \eprint{1704.04945}.

\bibitem[{\citenamefont{Barman et~al.}(2018)\citenamefont{Barman, Bhattacharya,
  and Zakeri}}]{Barman:2018esi}
\bibinfo{author}{\bibfnamefont{B.}~\bibnamefont{Barman}},
  \bibinfo{author}{\bibfnamefont{S.}~\bibnamefont{Bhattacharya}},
  \bibnamefont{and} \bibinfo{author}{\bibfnamefont{M.}~\bibnamefont{Zakeri}},
  \bibinfo{journal}{JCAP} \textbf{\bibinfo{volume}{09}}, \bibinfo{pages}{023}
  (\bibinfo{year}{2018}), \eprint{1806.01129}.

\bibitem[{\citenamefont{Barman et~al.}(2020)\citenamefont{Barman, Bhattacharya,
  and Zakeri}}]{Barman:2019lvm}
\bibinfo{author}{\bibfnamefont{B.}~\bibnamefont{Barman}},
  \bibinfo{author}{\bibfnamefont{S.}~\bibnamefont{Bhattacharya}},
  \bibnamefont{and} \bibinfo{author}{\bibfnamefont{M.}~\bibnamefont{Zakeri}},
  \bibinfo{journal}{JCAP} \textbf{\bibinfo{volume}{02}}, \bibinfo{pages}{029}
  (\bibinfo{year}{2020}), \eprint{1905.07236}.

\bibitem[{\citenamefont{Abe et~al.}(2020)\citenamefont{Abe, Fujiwara, Hisano,
  and Matsushita}}]{Abe:2020mph}
\bibinfo{author}{\bibfnamefont{T.}~\bibnamefont{Abe}},
  \bibinfo{author}{\bibfnamefont{M.}~\bibnamefont{Fujiwara}},
  \bibinfo{author}{\bibfnamefont{J.}~\bibnamefont{Hisano}}, \bibnamefont{and}
  \bibinfo{author}{\bibfnamefont{K.}~\bibnamefont{Matsushita}},
  \bibinfo{journal}{JHEP} \textbf{\bibinfo{volume}{07}}, \bibinfo{pages}{136}
  (\bibinfo{year}{2020}), \eprint{2004.00884}.

\bibitem[{\citenamefont{Nomura et~al.}(2021)\citenamefont{Nomura, Okada, and
  Yun}}]{Nomura:2020zlm}
\bibinfo{author}{\bibfnamefont{T.}~\bibnamefont{Nomura}},
  \bibinfo{author}{\bibfnamefont{H.}~\bibnamefont{Okada}}, \bibnamefont{and}
  \bibinfo{author}{\bibfnamefont{S.}~\bibnamefont{Yun}},
  \bibinfo{journal}{JHEP} \textbf{\bibinfo{volume}{06}}, \bibinfo{pages}{122}
  (\bibinfo{year}{2021}), \eprint{2012.11377}.

\bibitem[{\citenamefont{Chowdhury and Saad}(2021)}]{Chowdhury:2021tnm}
\bibinfo{author}{\bibfnamefont{T.~A.} \bibnamefont{Chowdhury}}
  \bibnamefont{and} \bibinfo{author}{\bibfnamefont{S.}~\bibnamefont{Saad}}
  (\bibinfo{year}{2021}), \eprint{2107.11863}.

\end{thebibliography}

\end{document}